\documentclass[]{elsarticle}

\usepackage[T1]{fontenc} 
\usepackage[version=3]{mhchem} 
\usepackage{natbib}
\usepackage{url}
\usepackage{doi}
\usepackage{color}
\usepackage{latexsym} 
\usepackage{marginnote}
\usepackage{datetime}
\usepackage[version=3]{mhchem}
\usepackage{wrapfig}

\newcommand{\textsup}[1]{$^{#1}$}

\def\um{\,$\mu$m}
\def\degC{$^\circ$C} 
\def\Dprime{D\textsup{\prime}}

\begin{document}

\begin{frontmatter}

\title{Graphene Grown on Ge(001) from Atomic Source}

\author[ihp]{Gunther Lippert\corref{cor}}
\ead{lippert@ihp-microelectronics.com}

\author[ihp]{Jarek D\k{a}browski}
\author[ihp]{Thomas Schroeder}
\author[ihp]{Yuji Yamamoto}
\author{Felix Herziger}%

\author[tub]{Janina Maultzsch}%

\author[luh]{Jens Baringhaus}
\author[luh]{Christoph Tegenkamp}

\author[soleil]{Maria Carmen Asensio}
\author[soleil]{Jose Avila}

\author[ihp]{Grzegorz Lupina}

\cortext[cor]{Corresponding author}

\address[ihp]{IHP, Im Technologiepark 25, 15236 Frankfurt (Oder), Germany}
\address[tub]{Institut f\"ur Festk\"orperphysik, Technische Universit\"at Berlin, Hardenbergstr. 36, 10623 Berlin, Germany}
\address[luh]{Institut f\"ur Festk\"orperphysik, Leibniz Universit\"at Hannover,  Appelstr. 2, 30167 Hannover, Germany}
\address[soleil]{Synchrotron SOLEIL, Saint Aubin, BP 48 91192 Gif-sur-Yvette, France}

\begin{abstract}
Among the many anticipated applications of graphene, some -- such as transistors for Si microelectronics -- would greatly benefit from the possibility to deposit graphene directly on a semiconductor grown on a Si wafer. 
We report that Ge(001) layers on Si(001) wafers can be uniformly covered with graphene at temperatures between $800^\circ$C and the melting temperature of Ge. The graphene is closed, with sheet resistivity strongly decreasing with growth temperature, weakly decreasing with the amount of deposited C, and reaching down to 2\,k$\Omega/\Box$. Activation energy of surface roughness is low (about 0.66\,eV) and constant throughout the range of temperatures in which graphene is formed. Density functional theory calculations indicate that the major physical processes affecting the growth are: (1)~substitution of Ge in surface dimers by C, (2)~interaction between C clusters and Ge monomers, and (3)~formation of chemical bonds between graphene edge and Ge(001), and that the processes 1 and 2 are surpassed by \ce{CH2} surface diffusion when the C atoms are delivered from \ce{CH4}. 
The results of this study indicate that graphene can be produced directly at the active region of the transistor in a process compatible with the Si technology.
\end{abstract}

\end{frontmatter}


\section{Introduction}

A key requirement for the realization of the variety of envisioned graphene applications \cite{Grapheneroadmap2012} is the availability of production methods delivering material with quality tailored to the specific needs of the particular application \cite{reviewferrari2012}. Microelectronics will likely require the highest quality graphene deposited inexpensively on large areas. Furthermore, graphene electronics with its anticipated unique features will most probably not be a stand-alone technology but will complement the existing technologies with new functionality. The ideal graphene deposition method should thus be compatible with the mainstream Si technology requirements and allow to grow high quality graphene directly on CMOS compatible dielectric and semiconducting substrates \cite{samsungnature11}. 

Currently, large area graphene \cite{ruoff2012Cu} and even heterostructures of 2D materials including graphene, \ce{MoS2} and BN \cite{ruoffCVDGraphonCVDBN, MOS2Grap2012} can be grown by CVD with high quality on metals such as Cu or Ni. Fabrication of electronic devices requires subsequent transfer to the target substrate. A variety of graphene devices can be then produced, as planar field-effect transistors \cite{IBM2012}, vertical transistors \cite{Sam2012Nanolett}, and Schottky diodes \cite{tongayPRX2012}, to name a few. Although transfer of graphene may be a viable option in some applications, it is not a generally preferred solution in microelectronic manufacturing where direct deposition would be ideal \cite{reviewferrari2012}.

Direct growth on Si has been studied with not necessarily encouraging results \cite{hackley,maeda} due to the high reactivity of Si against C, resulting in the formation of SiC \cite{CSiPhasediag}. Germanium does not form a stable carbide; the Ge-C and Cu-C systems\cite{CGePhasediag,CCuPhasediag} are similar. Ge is a semiconductor compatible with the Si technology and graphene grown on it can be directly used in such devices as the graphene base transistor \cite{Sam2012Nanolett,Mehr2012,GBT_Lecce2013}. Demonstration of catalyst-mediated \cite{sutter2006Ge} and catalyst-free \cite{angewandt2013} growth of graphene on Ge nanowires, and notably from \ce{CH4} on Ge \cite{Wang2013srep} make Ge a promising substrate. Yet, the growth of good graphene from \ce{CH4} requires much higher temperatures on Ge \cite{Wang2013srep} than on hexagonal BN \cite{GonBN_Yang2013nmat}, indicating that the Ge-C interaction plays a major role. This can be directly addressed by using atomic C instead of \ce{CH4}. 

We report on the first such study for this system. We show that graphene can be grown from atomic beam on Ge(001)/Si(001) and we use ab initio theory to analyze the C-Ge interaction with and without the presence of hydrogen. In accordance with the results of the CVD study \cite{Wang2013srep}, we find that the quality of graphene visibly improves if the Ge layer begins to melt during the deposition \cite{Ge001_surfaceMeltSantoni2003}; also this observation highlights the importance of the C-Ge interaction for the growth process. The unwanted side effect of the melting is however long-range roughening of the substrate. Given that the anticipated use of graphene grown on Ge(001) is in vertical transistor structures, in which the carriers travel across the interface between graphene and the germanium layer, such roughening is awaited to be at least problematic. Furthermore, heating the Ge layer up to temperatures close to the melting point is nearly certain to destroy any dopant profile in the layer. Studies of C-Ge interaction, as the study that constitutes a part of this work, may advance the knowledge needed to lower the growth temperature into the regime of safely low temperatures.

\section{Results and Discussion}

\begin{figure}[th]
\mbox{}\hfill
\includegraphics[width=0.95\textwidth,clip, trim=0mm 0mm 0mm 0mm]{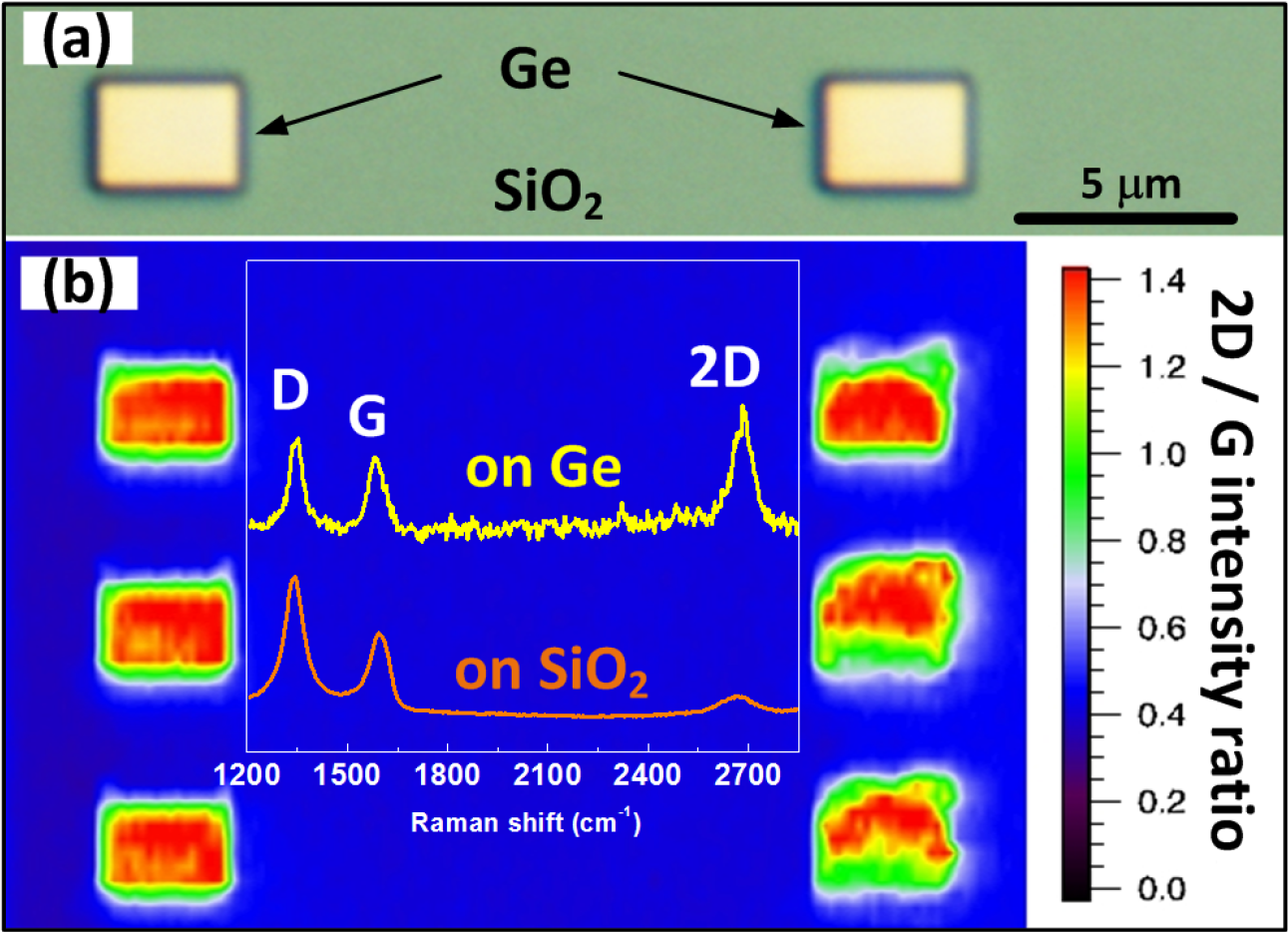}
\hfill\mbox{}
\caption{\label{fig:MT_selectivity} The difference in quality of graphene produced on Ge and on SiO$_{2}$. (a) Optical microscope image of Ge pillars embedded in SiO$_{2}$ matrix. (b) Raman spectra and 2D/G mode intensity map. In all cases investigated in this study, if the substrate temperature $T_{\rm sub}$ exceeds about 750\degC, graphene deposited on Ge has much higher quality than deposited on SiO$_{2}$ at the same conditions (here, $T_{\rm sub}$ = 900\degC). } 
\end{figure}

We apply the molecular beam epitaxy (MBE) process used for the growth of graphene on van der Waals substrates \cite{Lippert2013Carbon,lippert} to deposit carbon atoms onto Ge(001)/Si(001) templates. Ge layers are grown on Si(100) in tailored multistep processes compatible with standard Si technology \cite{Yamamoto2012}. High uniformity, low threading dislocation density, and low surface roughness of Ge can be achieved, providing a high-quality substrate for graphene deposition (see Methods). An oxide-free Ge surface is prepared using a combination of wet-etching and UHV annealing and exposed to a beam of thermally evaporated carbon atoms at various substrate temperatures. Analysis of the surface chemical composition and direct comparison with HF-last Si substrates reveals that in contrast to Si, Ge does not form a stable carbide phase (cf. Supporting Information for XPS spectra). Instead, as it is proved by Raman spectroscopy, at elevated temperatures the C deposit on Ge takes a form of sp$^2$-hybridized carbon layer. 

Experiments performed on SiO$_2$-patterned Ge substrates (Fig.\,\ref{fig:MT_selectivity}) demonstrate that graphene is produced (as visualized by the 2D/G intensity ratio) and that there is marked difference in the quality of graphene produced at the same conditions on both materials. As we explain in the course of the discussion, albeit the graphene film still contains numerous defects and/or the domain size is clearly smaller than that achievable by growth using chemical vapor deposition (CVD) from \ce{CH4} and \ce{H2} mixture at atmospheric pressure \citep{Wang2013srep}, the electrical properties of graphene obtained in the current study by MBE are good enough to qualify it for the use in a terahertz graphene base transistor. We also analyze the reason for the observed differences in the quality of graphene grown on Ge(0010) substrates by CVD and by MBE.

\begin{figure}[th]
\includegraphics[width=1.05\columnwidth]{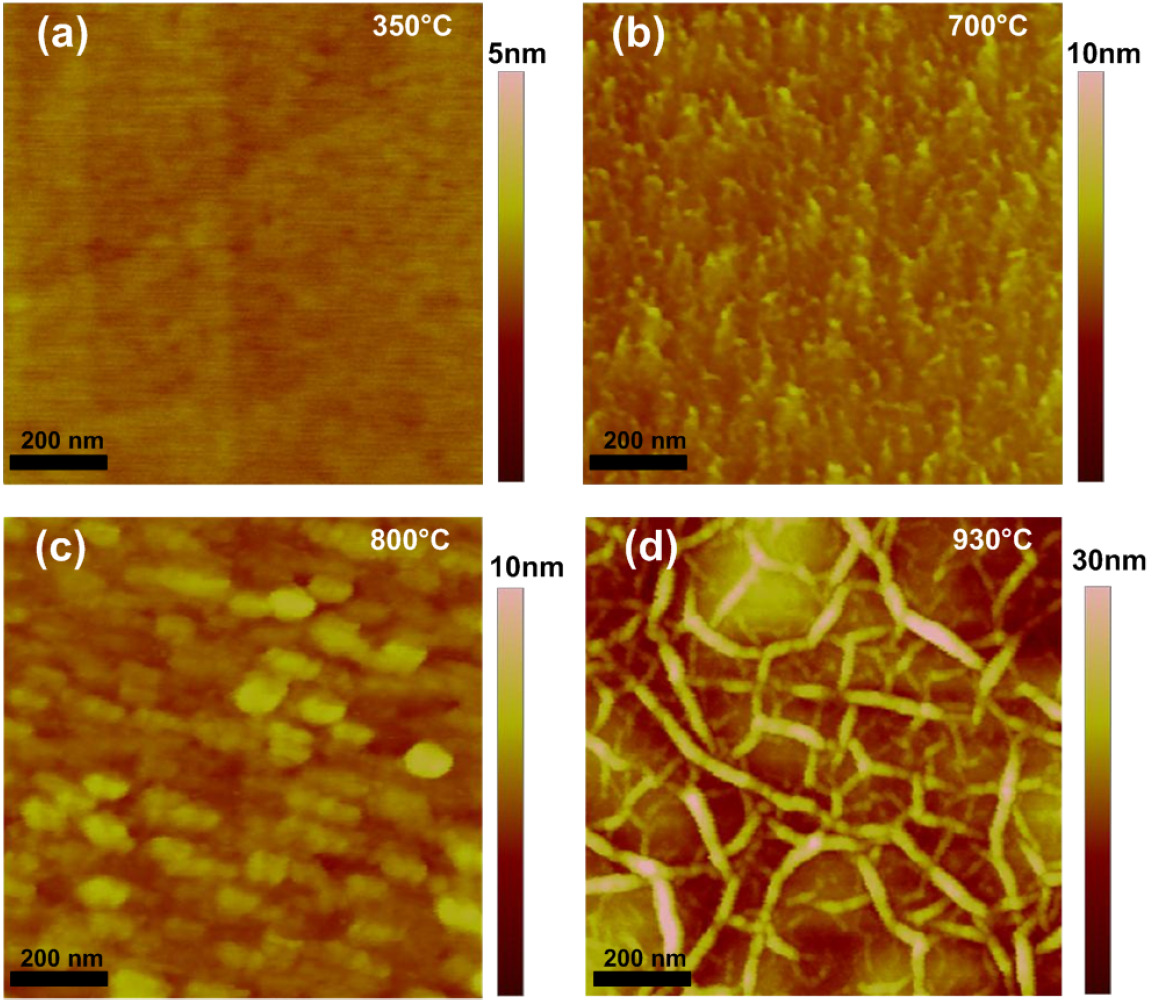}
\caption{\label{fig:afm} AFM images acquired after C deposition at various substrate temperatures.} 
\end{figure}

On Ge(001), the highest crystalline quality and lowest sheet resistance of the graphene layers is obtained at temperature approaching the melting point of the substrate. The drawback of high-temperature growth is an increased surface roughness, revealed by atomic force microscopy (AFM) images, cf. Fig.\,\ref{fig:afm}. Clearly, the surface topography changes significantly with increasing substrate temperature. While at low temperatures (below 550\degC) the surface roughness remains comparable with roughness of the initial Ge surface (0.12-0.16\,nm), higher substrate temperature during growth results in strongly increased root mean square (rms) roughness exceeding 1\,nm for growth temperature above 700\degC; furthermore, at 930\degC\ high-frequency wrinkles and low-frequency hills appear. The mechanisms responsible for this roughening are discussed by the end of this section on the basis of the measured activation energy and of the characteristic features of C-Ge interaction revealed by ab initio calculations.

\begin{figure}[th]
\mbox{}\hfill
\includegraphics[width=0.9\columnwidth,clip, trim=0mm 0mm 0mm 0mm]{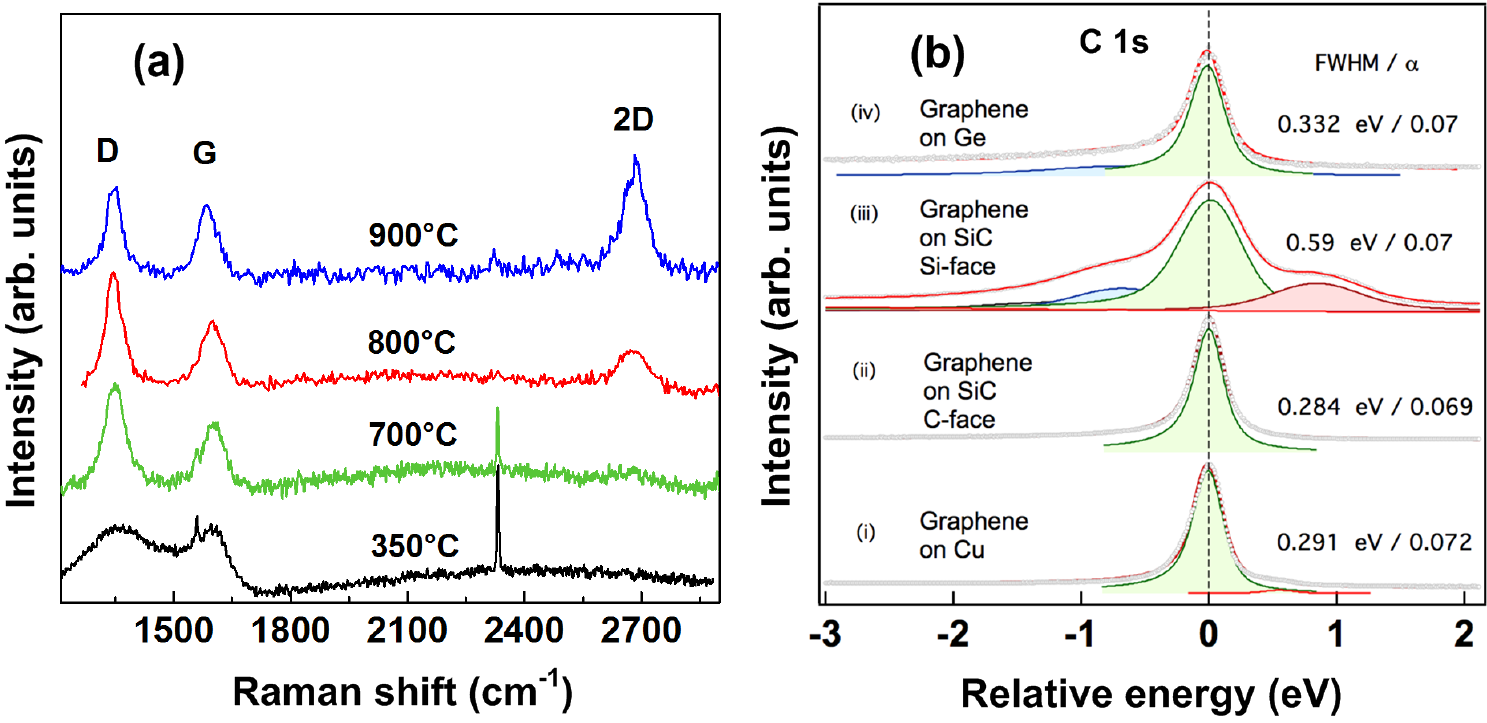}
\hfill\mbox{}
\caption{\label{fig:RamanSpectra} 
(a) Raman spectra of films grown on Ge at various substrate temperatures by deposition of about 5 carbon monolayers. %
The peaks 1555\,cm$^{-1}$ and 2350\,cm$^{-1}$ are attributed to atmospheric oxygen and nitrogen that appear due to long integration times.
(b)~High resolution synchrotron radiation XPS. C\,1s peak shape ($T_{\rm growth}$ = 850\degC), compared to typical shapes for graphene grown on other substrates. The information depth is about 2 monolayers. The green component comes from graphene, the blue one may be due to O contamination. See the Supporting Information for more discussion of XPS and Raman spectra.
} 
\end{figure}

Figure \ref{fig:RamanSpectra}a shows Raman spectra acquired from samples prepared at various substrate temperatures and at the same growth rate (estimated to be about 1.4 monolayers per minute) and time (200\,s). %
The 2D Raman peak typical for sp$^2$ carbon \cite{Tuinstra1970} can be resolved only in films grown above about 750\degC, but the 2D/G peak area ratio is then close to 1 already at 800$^\circ$C and approaches 2 above 900\degC, when the substrate begins to melt (Fig.\,\ref{fig:RamanSpectra}a). Such a high 2D/G ratio indicates that already at 800\degC\ sp$^2$-hybridized carbon, i.e. graphene, covers most of the surface. In the C 1s core level peak in the XPS spectrum (Fig.\,\ref{fig:RamanSpectra}b), no other bonds as C-C sp$^2$ can be clearly resolved. The FWHM reflects the degree of crystallinity and strain; it is only slightly larger than that for CVD graphene grown on copper. The parameter $\alpha$ reflects asymmetry; the value of $\alpha=0.07$ implies graphene structure and metallic conductivity (cf. the Supporting Information on XPS).

\begin{figure}[th]
\includegraphics[width=1.0\columnwidth,clip,trim=0mm 0mm 0mm 0mm]{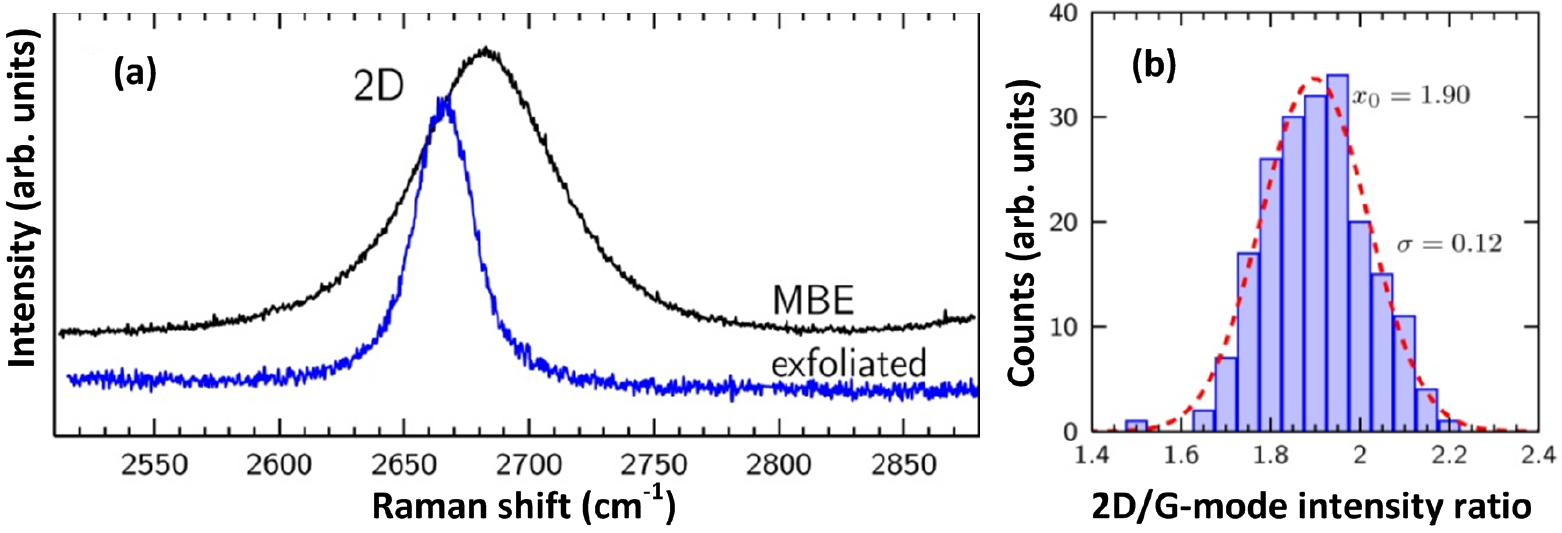}
\caption{\label{fig:raman_xy}%
(a) Raman spectra of the MBE graphene sample grown at 930\degC\ (black) and exfoliated single-layer graphene (blue) on germanium in the range of the 2D mode. 
(b) Analysis of the 2D/G-mode intensity ratio $r_{\rm 2D/G}$ in the investigated area (10$\times$30\,\um$^2$) of the same MBE sample. The intensity of the G and 2D mode is evaluated from the peak area. The data was fitted assuming a Gaussian distribution (red, dashed lines).
}
\end{figure}

The 2D Raman mode stems from an inter-valley double-resonant scattering process involving two TO phonons close to the K point (the Dirac point in single-layer graphene) on the Brillouin zone boundary. Since the double-resonance process depends on both the electronic band structure and the phonon dispersion \cite{thomsen2000}, the 2D-mode line shape gives information about both properties \cite{ferrari2006}. From the line shape (Fig\,\ref{fig:raman_xy}a) 
and relative intensity of the 2D peak we conclude that the graphene consists of decoupled layers and is not single-layer graphene. The 2D mode of the MBE-grown graphene is symmetric with a single-Lorentzian shape, but strongly broadened and up-shifted in comparison to that of graphene exfoliated on the same substrate; the same is true for the G peak (see the Supporting Information). Such behavior indicates the presence of nanocrystalline graphene.\cite{berciaud2009} The peak positions and widths (FWHM) of the G and 2D peaks may in principle be used to evaluate the doping level or the strain in graphene, but in this particular case they are more likely to be dominated by the nanocrystallinity of graphene. 


The statistical analysis of the G and 2D peak positions and the 2D/G intensity ratio (Fig.\,\ref{fig:raman_xy}b and Supporting Information) shows very low standard deviations (0.04\,cm$^{-1}$ and 0.12\,cm$^{-1}$, respectively), demonstrating the very homogeneous growth of graphene on Ge.

The average 2D/G intensity ratio of 1.9$\pm$0.12 is significantly lower than the typical value of four typical for exfoliated single-layer 
graphene\cite{ferrari2006}, which may again indicate that the film has multiple layers. However, graphene that is doped, interacts with the environment or is imperfect, may also have the ratio below four, even around one.\cite{Kalbac2010,Cheng2010}

\begin{figure}[th]
\mbox{}\hfill
\includegraphics[width=0.9\columnwidth,clip, trim=0mm 0mm 0mm 0mm]{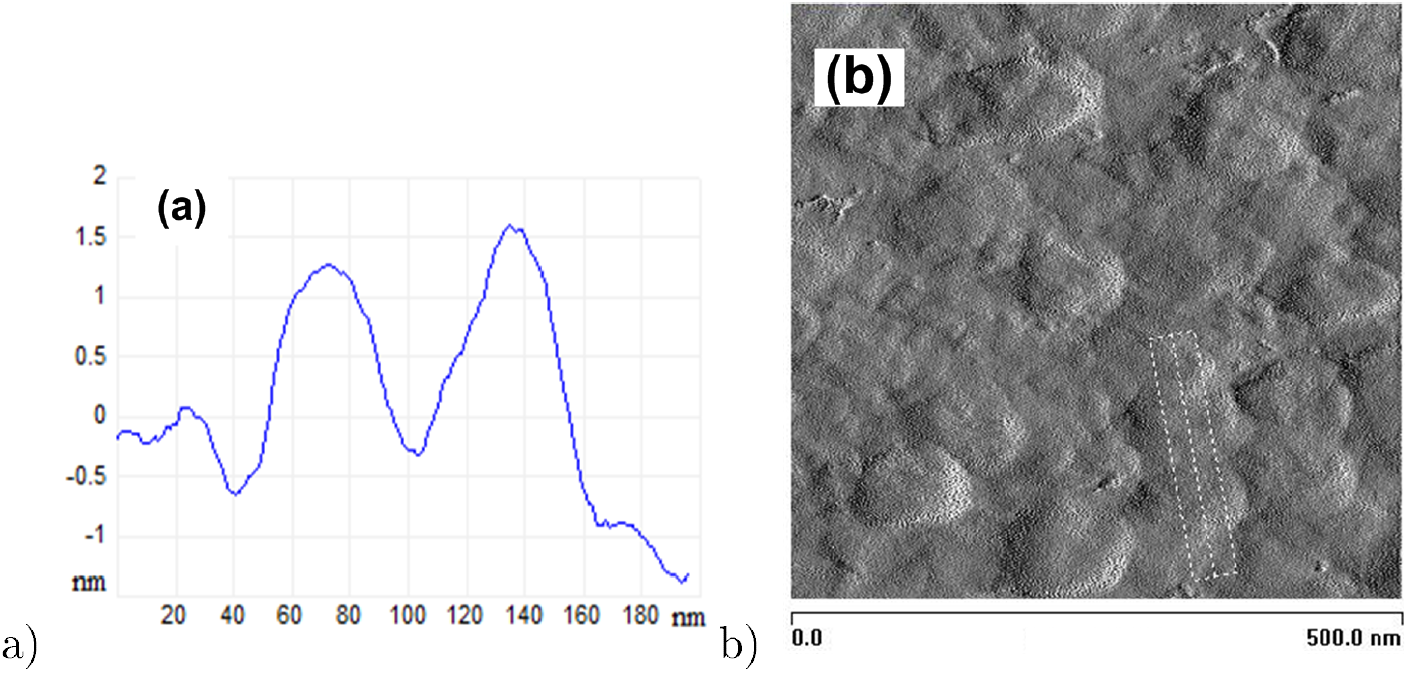}
\hfill\mbox{}
\caption{\label{fig:AFM_grains} Graphene grains in the sample on which about 5 monolayers of C have been deposited at 800\degC. %
(a) AFM corrugation averaged along the axis of the box shown in the panel b. (b) AFM amplitude image illustrating the sub-structure of the grains in the length scale below the grain size of about 20-50 nm recognizable in the AFM height scan (panel a). } 
\end{figure}

In contrast to the 2D Raman peak, the D mode scattering process involves a TO phonon and a defect. The D peak is absent in not sp$^2$-hybridized carbon and symmetry-forbidden in perfect graphene or graphite. Therefore, it is a measure of disorder  in the sp$^2$ carbon network. Its relative intensity reflects thus the density of defects, in particular, the presence of boundaries. The behavior of the D peak as revealed by Fig.\,\ref{fig:RamanSpectra}a shows that sp$^2$-bonded carbon is produced already at low substrate temperatures and that disorder is considerable also in samples grown at high temperatures. The  D/G intensity ratio can be used to estimate the grain size of nanocrystalline graphene \cite{Tuinstra1970,Cancado2007,Lucchese2010}; applying this method we deduce the average crystalline grain size to be about 10\,nm.  AFM topology is consistent with this estimate at least to the order of magnitude (Fig.\,\ref{fig:AFM_grains}). The AFM amplitude indicates that each of the grains has a sub-structure, which may explain the difference in these two estimates.

\begin{figure}[th]
\mbox{}\hfill
\includegraphics[width=0.9\columnwidth,clip, trim=0mm 0mm 0mm 0mm]{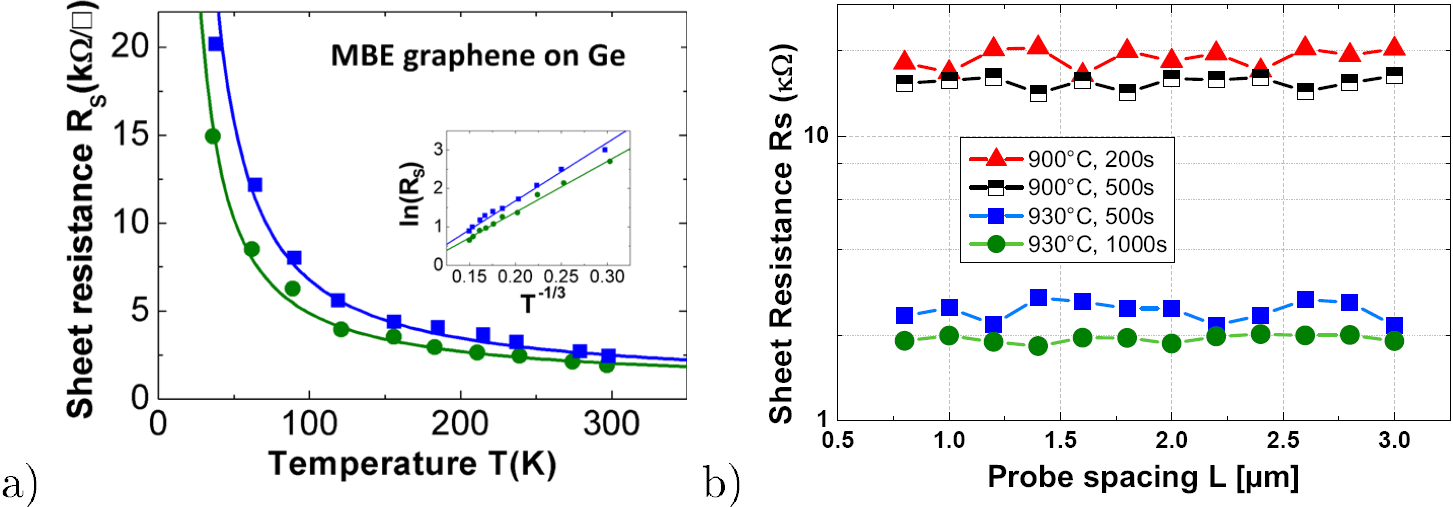}
\hfill\mbox{}
\caption{\label{fig:Electrical} 
(a) Sheet resistance measured by 4-point STM as a function of temperature for a fixed probe spacing $L = 2$\,\um\ (cf. Supporting Information for technical details). The symbols are the same as in panel b. 
(b) Sheet resistance as a function of probe spacing. 
} 
\end{figure}

Electrical measurements performed with 4 colinearly arranged STM tips (see Supporting Information for detailed discussion) reveal ohmic IV characteristics of the MBE graphene. The metallic character is retained at temperatures below which the Ge(001) surface is semiconducting, i.e., below 200\,K \cite{Jeon2006}; this excludes any contribution from substrate to the measured currents. However, the film contains high density of defects causing disorder in the electrostatic potential. This follows from the results of temperature dependent measurements (Fig.\,\ref{fig:Electrical}a). With decreasing sample temperature the sheet resistance exponentially increases from 2\,k$\Omega/\Box$ up to 20\,k$\Omega/\Box$. Such a behavior is well known for disordered systems, including disordered graphene. It can be understood in terms of Anderson localization and variable range hopping (VRH) transport \cite{Anderson1958,Peters2012,Yan2010,Cheah2013,Mott1971}. The fit in Fig.\,\ref{fig:Electrical}a proves that the dependence of $\log(R_{\rm s})$ on $T^{-1/3}$ is linear, which is a signature of two-dimensional VRH. Indeed, as obvious from Fig. \ref{fig:Electrical}b) the resistivity is independent of the probe spacing clearly indicating 2D transport.

As the substrate temperature during growth is increased from 900\degC\ to 930\degC, the sheet resistivity $R_{\rm s}$ (Fig.\,\ref{fig:Electrical}b) drops fast to 2\,k$\Omega/\Box$, a value comparable to $R_{\rm s}$ of CVD graphene\cite{sheetResistanceLi2009,sheetResistanceGunes2010,sheetResistanceBiswas2011,sheetResistanceIHPremark} or to typical $R_{\rm s}$ of base layer in a SiGe HBT transistor. On the other hand, there is little dependence of $R_{\rm s}$ on the amount of deposited C. This may indicate that only the topmost carbon layers contribute significantly to the electrical conductivity of the film, or that the carbon that is deposited above a certain critical amount agglomerates in grains.

\begin{figure}[th]
\mbox{}\hfill
\includegraphics[width=0.95\columnwidth,clip, trim=0mm 0mm 0mm 0mm]{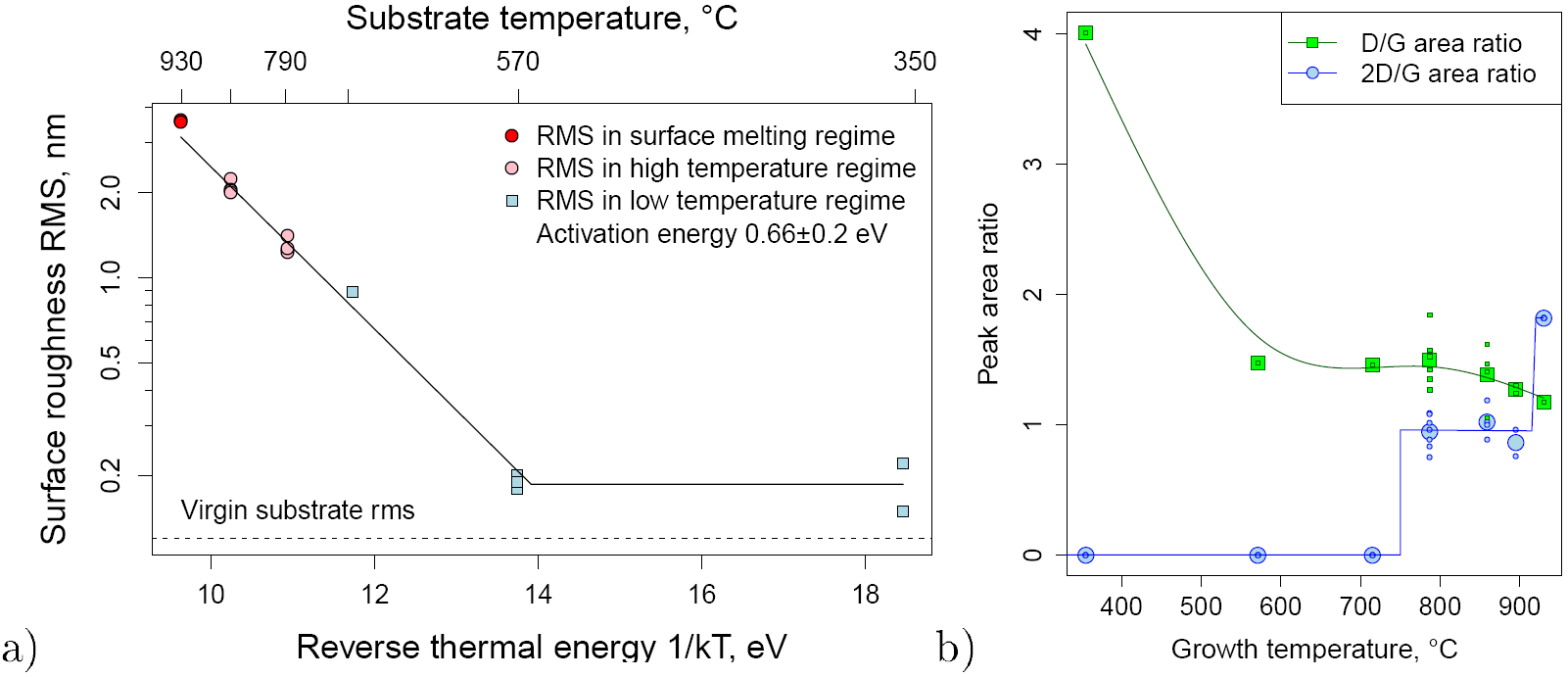}
\hfill\mbox{}
\caption{\label{fig:RMS}
(a) Arrhenius plot of AFM surface roughness RMS of nominally 6-monolayer films.
(b) D/G and 2D/G Raman peak area ratio dependence on the growth temperature. 
} 
\end{figure}

AFM statistics provides more support to the second of these hypotheses. Strong height variations appear in AFM topography on sub-micrometer length scale when the temperature exceeds about 600$^\circ$C (Fig. \ref{fig:RMS}a): the film consists of grains. The surface roughness rms monotonously increases with the substrate temperature, suggesting that the roughening occurs by Ostwald ripening.\cite{OswaldLifshitz1961} 
Ostwald ripening of grains is a phenomenon natural to expect during growth and for the purpose of further discussion we assume that this the mechanism that controls the grain evolution. 

When the substrate begins to melt, the surface topology changes: instead of randomly distributed grains, wrinkles forming a network of lines appear (Fig.\,\ref{fig:afm}d). At the same time, the resistivity of graphene drops and the 2D/G Raman mode ratio sharply increases (Fig.\,\ref{fig:RMS}b). The activation energy $E_{\rm RMS}$ of the surface roughness rms, as obtained from Arrhenius plot, remains however the same in the whole range of temperatures (Fig. \ref{fig:RMS}a). It is the same not only below 750\degC, where no 2D peak can be resolved (Fig.\,\ref{fig:RamanSpectra}a and \ref{fig:RMS}b), and between 750\degC\ and 900\degC, where the 2D/G ratio $r_{\rm 2D/G}$ remains close to 1, but also in the melting regime, where $r_{\rm 2D/G}$ rises to 2. It seems that around 750\degC\ a nanocrystalline graphene layer is formed on top of the film, and that this layer improves as the substrate begins to melt.

The growth of grains supporting the graphene layer is limited by a process with low barrier height of $E_{\rm RMS}$ = 0.66\,eV, comparable to the barrier for migration of C ad-atom on graphene \cite{Nieminen2003}. This implies that C atoms are easily liberated from one grain and then easily diffuse to another grain. Such easy detachment is hard to understand if the grain boundaries consisted of pure C, but can be rationalized if they are contaminated with Ge: as will be shown from DFT calculations, the barrier for a process in which Ge and C atoms exchange places may be as low as 0.65\,eV, at least when the Ge atom belongs to Ge(001).

\begin{figure}[th]
\includegraphics[width=0.95\columnwidth,clip, trim=0mm 0mm 0mm 0mm]{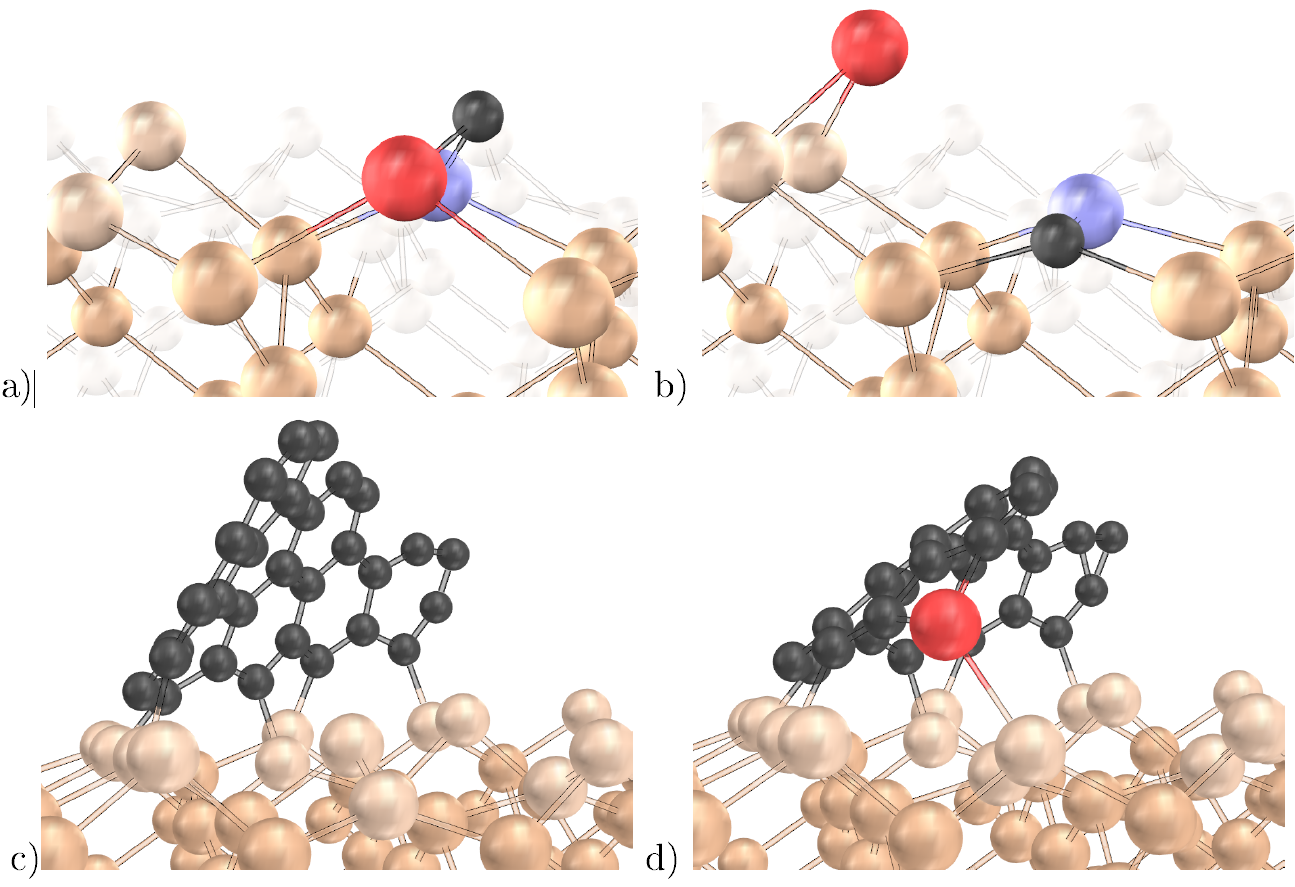}
\caption{\label{fig:Islands}
(a)~Carbon atom (black) adsorbed on Ge(001) dimer (red and blue atoms). 
(b)~C atom substitutes Ge in the dimer and ejects a Ge atom.
(c)~A piece of graphene chemically bonded to Ge surface. 
(d)~ The ejected Ge atom (red, cf. panel c) attaches itself to its edge.
} 
\end{figure}

\begin{figure}[th]
\includegraphics[width=1\columnwidth,clip, trim=0mm 0mm 0mm 0mm]{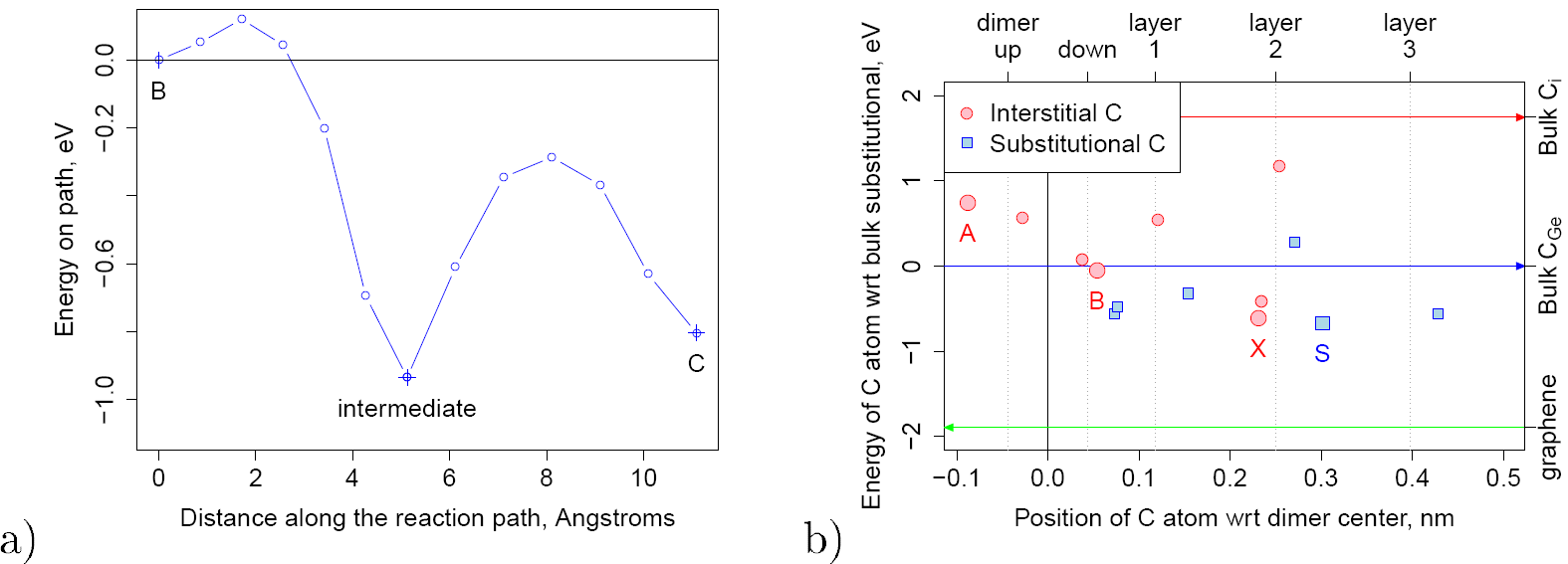}
\caption{\label{fig:Ejection_and_DefectEnergies} 
a) Energy barriers for the process of Ge ejection. Adsorption of atom from vacuum to the configuration "a" proceeds withe the energy gain of 4.6\,eV and no barrier. The states "A" and "B" are shown in Fig. \ref{fig:Islands}a-b. Carbon in the intermediate state sits directly under the Ge dimer; the dimer bond is broken. 
b) Energy of C atom in various configurations, measured with respect to the energy of the  substitutional C in the bulk. The ordinate is the vertical distance to the center of Ge dimers; the positions of dimer atoms and atomic layers of the perfect surface are indicated by dotted lines. Labeled larger symbols correspond to the configurations discussed: "A" and "B" are the structures shown in Fig. \ref{fig:Islands}a and \ref{fig:Islands}b, "X" is the lowest-energy subsurface interstitial, and "S" is the lowest-energy subsurface substitutional.
}\end{figure}

Also the very presence of Ge at the boundaries can be understood on the basis of ab initio calculations (Fig.\,\ref{fig:Islands}, cf. also the Supporting Information). The interaction of C atoms with the reconstructed Ge(001) surface leads to ejection of Ge atoms. As illustrated in Fig\,\ref{fig:Islands}a-b, a C atom readily substitutes a surface Ge atom, which is kicked out into a mobile on-surface state. The energy barrier for ejection is only about 0.65\,eV (Fig.\,\ref{fig:Ejection_and_DefectEnergies}a), low enough for the process to take place rapidly at any deposition temperature used in this experimental study. 

The ejected Ge atoms become ad-atoms and are highly mobile. Their total amount is expected to be comparable to a monolayer. This is suggested by DFT molecular dynamics, according to which the probability of ejection before the adsorbed C atom thermalizes is of the order of 50\%. The thermalized C atom resides for a while (microseconds at 850\degC, milliseconds at 450\degC) directly under the surface, as interstitials. Being strongly repelled from the bulk by elastic forces (Fig.\,\ref{fig:Ejection_and_DefectEnergies}b), it remains close to the surface and eventually ejects a Ge atom (Fig.\,\ref{fig:Islands}b) or diffuses under the surface and then attaches itself to a C cluster (Fig.\,\ref{fig:Islands}c-d). The probability that a non-surface Ge atom will be substituted is negligible. As for the ejected Ge atoms, also they are trapped by the clusters (Fig.\,\ref{fig:Islands}d), unless they find their way to a surface step or liberate a C atom from a C-Ge dimer by a kick-in process (reverse to Ge kick-out shown in Fig.\,\ref{fig:Ejection_and_DefectEnergies}a).

Indeed, the ejected Ge atoms are preferably attached between graphene edge and Ge(001). Depending on the adsorption geometry, the energy of the attached atom is by 0.6\,eV to 2.3\,eV lower than in the bulk. Furthermore, when a piece of graphene is placed on Ge(001), its edge atoms tend to make bonds with the substrate. Figure \ref{fig:Islands}c-d illustrates both types of the graphene-Ge interaction for the case of a \ce{C29} molecule. The molecule attached itself to Ge(001) with five of its edge atoms (Fig.\,\ref{fig:Islands}c); graphene-substrate bond dissociation energy was 1.5\,eV per C-Ge bond. In its stable state the molecule is bonded along the whole edge, as predicted for graphene nanoribbons on Si(001) \cite{nanoribbon_Zhang2009}, and the bond dissociation energy is 1.0\,eV per C-Ge bond. But this optimum is difficult to reach for all nucleated graphene pieces (see the discussion in Supporting Information) and many of the molecules are expected to be trapped in a metastable state similar to that depicted in Fig.\,\ref{fig:Islands}c. When a Ge ad-atom attaches itself between this molecule and the substrate (Fig.\,\ref{fig:Islands}d) the energy is lowered by 2.8\,eV and the molecule tilts back towards the horizontal orientation. Both tendencies (to stand up and to trap Ge) are pronounced, hence they both should have noticeable influence on the growth mode. 
Yet this influence is likely to be smaller when C is delivered from a molecular beam (as in this study) than when it is delivered from a mixture of \ce{CH4} and \ce{H2} (as in the CVD study reported in Ref.\,\cite{Wang2013srep}), because hydrogen should preferentially etch graphene at sites where the edge is not protected by bonds formed with Ge(001), thus reducing the need to "glue" the standing molecules back to the substrate. Indeed, the CVD films are markedly better (in terms of D Raman mode intensity) than the MBE films grown at the same substrate temperature.

We suppose that the deterioration of the graphene growth mode by the formation of graphene-substrate bonds is reduced in MBE by the supply of C from subsurface "X" interstitial sites. These atoms have easier access to the graphene edge that is bonded to the substrate than to free parts of the edge.

The dominating mechanism by which carbon is supplied to the growing graphene is affected by the presence of hydrogen. Albeit -- according to DFT results -- when \ce{CH4} molecule comes into a chemical contact with the surface, it adsorbs dissociatively, losing the first two hydrogen atoms one by one with  barrier low enough to play no decisive role at the optimal deposition temperatures, the \ce{CH2} produced by this reaction sequence is a relatively stable species on Ge(001). Carbon deposited from \ce{CH4} is estimated to diffuse on the surface in the form of \ce{CH2}, with the barrier of about 1.5\,eV. The \ce{CH2} molecule diffuses for a longer while (from hundreds miliseconds at 900\degC\ to about a second at 800\degC) before it decays into a subsurface "X" carbon and hydrogen atoms terminating the surface dimer atoms. The distance covered during this time at these temperatures is of the order of 200\,nm and the life time of \ce{CH2} is long enough (by the \ce{CH4} flow rates used for deposition) for the molecule to encounter another molecule before decay takes place. In the CVD process, most of C atoms are therefore expected to be delivered to graphene from the top of the surface, while in the MBE process they should arrive from under the surface. For the same reason, the amount of ejected Ge should be considerably smaller during CVD than during MBE.

It follows that by varying the carbon flux (carbon source temperature in MBE, \ce{CH4} flow in CVD) and the availability of \ce{H2}, one tunes the balance between all these carbon delivery processes and influences the chemistry of at the edge of the growing graphene. 

The surface roughening observed in the MBE process (and not reported for the CVD process) may be associated with the tendency of small graphene molecules to stand up. Since the boundaries of grains forming the roughened film are expected to be decorated with Ge atoms, it is plausible that on the grain boundaries there exist edge sites, from which atoms or dimers detach with energy barrier compatible to that observed in the experiment (Fig.\,\ref{fig:RMS}a). One may speculate that the reaction that forms the bottleneck in the Ostwald ripening of the grains and is characterized by the measured barrier height  $E_{\rm RMS}$ = 0.66\,eV is associated with Ge-mediated migration of grain boundary planes. The grains would then consist of graphene stacks with edge partially glued to Ge(001) by ejected Ge atoms, and partially glued to other such stacks, again by Ge atoms.

Finally we note that the observed defected character of the MBE film (relatively strong D Raman mode) is not associated with defects in the virgin Ge layers, because the defect density (dislocation density) in the Ge layer \cite{Yamamoto2012} corresponds to the average defect-defect distance of several \um, which is well below the average domain size deduced from the Raman spectra of the MBE graphene. The defects giving rise to the D Raman mode are not associated also with diffusion of Si from the wafer to the Ge(001) surface, because the amount of such Si should increase with the substrate temperature, while in contrary to this the quality of the film improves with the substrate temperature, being clearly the best when the Ge layer begins to melt. However, one cannot fully exclude the possibility that the observed roughening of the graphene film is connected with contamination of the surface with Si atoms segregated from the wafer. Yet if the Si contamination during the growth would be responsible for the roughening, one would expect an activation energy reflecting the activation energy of Si diffusion in Ge. The rms surface roughness should be thus activated with nearly 3\,eV \cite{SiinGeDiffusionRaisanen1891}. The measured activation energy of 0.66\,eV is  too low to account for this process. We therefore tend to associate the roughening of the graphene film with the supposed vertical orientation of some of the graphene nuclei.

\section{Conclusions}

\begin{figure}[th]
\mbox{}\hfill
\includegraphics[width=0.75\columnwidth,clip, trim=0mm 0mm 0mm 0mm]{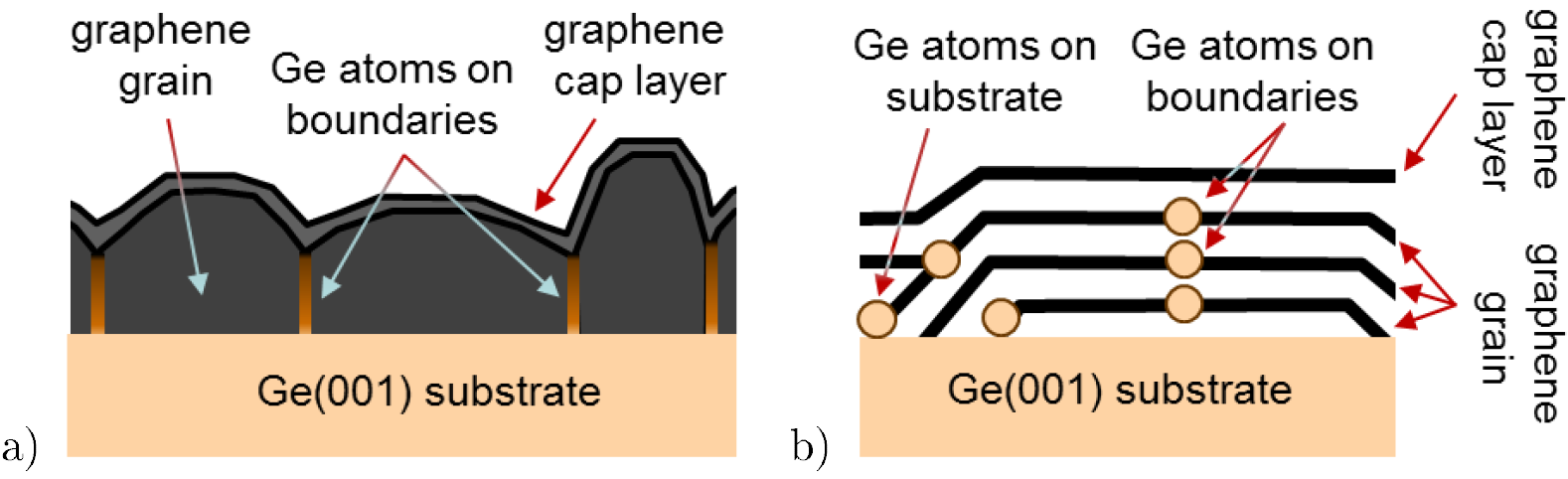}
\hfill\mbox{}
\caption{\label{fig:Model_grains} 
Supposed structure of MBE graphene on Ge(001).
a) Overview. b) Details. 
}\end{figure}

Germanium does not mix with carbon and as such it is a suitable substrate to grow graphene. We have demonstrated that molecular beam growth can be used to uniformly cover with decoupled few-layer graphene a Ge(001) film grown on a Si(001) wafer. The graphene sheets are free of carbon nanotubes and consist predominantly of grains with diameter in the range of tens of nanometers. This can be achieved at temperatures between $800^\circ$C and the melting temperature of Ge. Films deposited at lower substrate temperatures are uniform as well and consist of sp$^2$-bonded carbon (G Raman mode), but have no Raman signature of ordered sixfold rings (2D mode). On the basis of Raman and XPS spectra, AFM measurements, electrical measurements, and ab initio DFT calculations we argued that films deposited above $750^\circ$C consist of grains built of stacked graphene layers (Fig.\,\ref{fig:Model_grains}b).
The measured low activation energy of the surface roughness rms (0.66\,eV) suggests that the grain facets are contaminated with Ge. This is compatible with the DFT prediction that interaction of atomic C with the clean Ge(001) leads to ejection of surface Ge atoms to mobile ad-atom (monomer) state and that these Ge monomers are preferentially adsorbed on graphene edges.

The grains of graphene seem to have nucleated in the initial stage of growth, when small graphene molecules have a tendency to stand up vertically because their edges make chemical bonds with the substrate. We argue that this tendency is reduced by Ge atoms released from the substrate by C atoms (Fig.\,\ref{fig:Model_grains}b). Since the sheet resistance of the film hardly depends on film thickness and since the 2D/G Raman mode intensity ratio $r_{\rm 2D/G}$ depends step-wise on deposition temperature (no 2D below 750\degC, $r_{\rm 2D/G}\simeq1$ between 750\degC\ and 900\degC, and $r_{\rm 2D/G}\simeq2$ above 930\degC), we suppose that the grains are covered by a graphene film of higher quality (cf. the cap layer in Fig.\,\ref{fig:Model_grains}) that significantly improves when the attachment of graphene edges to Ge(001) weakens as the substrate begins to melt.

Indeed, AFM images show that as the substrate temperature approaches the melting point of germanium ($T^{\rm Ge}_{\rm melt}$ = 937\degC), the grains of similar height become arranged into interwoven lines, but the grain growth mechanism (as characterized by surface roughness RMS activation energy) remains the same as at lower substrate temperatures. The process of line network formation may be driven by a stress relaxation mechanism. This relaxation may lead to electrical and vibrational improvement of the graphene cap layer.

Electrical measurements with 4-point probe STM prove ohmic behavior and 2D conductivity of the film. Independently, metallic behavior follows also from the asymmetry of C\,1s core level XPS peak. Temperature dependence of the sheet resistance reveals the Anderson localization phenomenon, known to occur in disordered graphene. 

The sheet resistivity drops significantly, down to 2\,k$\Omega/\Box$ for samples grown at 930\degC. Such a low value is comparable to (albeit higher than) that achievable in standard CVD graphene grown on copper substrates \cite{sheetResistanceLi2009,sheetResistanceGunes2010,sheetResistanceBiswas2011,sheetResistanceIHPremark} and sufficient for application of the MBE graphene/Ge(001) film in a high-frequency graphene-base transistor. Transistors of this kind can be produced with graphene grown directly on the Ge(001) layer: there is no need to remove or further process the germanium.

These results indicate that germanium has some potential as a substrate for growth of graphene. The most advantageous property of germanium is here that, albeit it is a semiconductor, it does not form carbides and carbon hardly dissolves in it. The disadvantage is that carbon atoms at graphene molecule edges make chemical bonds with surface atoms of Ge(001). Nevertheless, Ge atoms ejected from the substrate act as glue that reduces the detrimental tendency of graphene molecules to stand up vertically on the surface and that works even at relatively low deposition temperatures. This leaves room for improvement of the film quality by properly tailored sequence of growth steps. 

The implications of this study are strengthened by the very recent report that atmospheric pressure CVD can be used to grow high-quality graphene on germanium wafers \cite{Wang2013srep}. Still, even the CVD process requires the substrate temperature to be above 900\degC, i.e., close to the the melting point of germanium. This is problematic, given that the direct growth of graphene on germanium is needed for devices such as the graphene base transistor, in which the transport of electrons takes place along the surface normal, that is, also from the graphene to the germanium substrate. When the substrate melts, the surface roughens (cf. the hills visible in Fig.\,\ref{fig:AFM_grains}d and forming a low-frequency pattern, independent of the high-frequency wrinkles), which may deteriorate the graphene-substrate interface by producing regions where the graphene hangs over above the substrate (i.e., the graphene-substrate distance is there significantly larger than when graphene is placed on top of a flat germanium surface). Other problems, such as strong broadening of the dopant profile by diffusion of dopant atoms or thickness inhomogeneity of the undoped germanium that should separate the graphene from doped germanium in such devices are anticipated as well. Lowering of the growth temperature seems therefore desirable.

From the comparison of the MBE and CVD results and from the accompanying DFT calculations one can conclude that in both cases the temperature around the surface melting point $T_{\rm sm}$ of germanium is needed as a consequence of carbon-germanium interaction at the graphene edge. It follows that in order to lower this temperature to under $T_{\rm sm}$ one should  focus further studies on the control of this chemistry. According to the analysis presented in this report, hydrogen affects the balance between the major mechanisms by which C and Ge atoms are delivered to the graphene edge and thus allows one to use its availability as a means to tune the growth process. The exact method to lower the growth temperature remains however to be found. 


\section{Methods}

\textbf{Deposition.} High quality Ge (001) layers used as substrates for graphene growth were deposited on non-patterned and patterned 200\,mm Si(001) wafers using reduced pressure chemical vapor deposition in a two-step process described in detail elsewhere \cite{Yamamoto2011,Yamamoto2012}. The thickness of Ge layer on non-patterned and patterned substrates was 1.1\,\um\ and 200\,nm, respectively. Clean Ge surfaces were prepared by dipping in HF:H$_2$O solution followed by a flash anneal at 760\degC\ for 60\,s \cite{Klesse2011Capellini}. 

The deposition of carbon was carried out in a DCA molecular beam ultra high vacuum (UHV) system on 1\,\um\ Ge(001) CVD films grown on Si(001) wafers. The growth rate was about 1.4 graphene monolayer per minute, as estimated from X-ray reflectivity (XRR) measurements on a thick (15\,nm) film grown at low temperature (200\degC) to suppress surface roughening (the sticking coefficient of C on graphite is close to 1 and its temperature dependence is weak).\cite{ConGstickingPhilips} The source (high-purity pyrolytic carbon) was placed 35 cm away from the sample and emitted mostly carbon atoms. The substrate temperature was varied between 350\degC\ and 930\degC. The growth time was typically 200\,s, with some attempts performed for 100\,s, 500\,s, and 1000\,s. The residual pressure during growth was in the range of 10\textsup{-7~}mbar. 

\textbf{Characterization.} The quality of the graphene film was studied \emph{ex-situ} by $\mu$-Raman spectroscopy using Renishaw In-viaFlex spectrometer and the green laser light ($\lambda$ = 514\,nm). Spatial resolution was 0.4\um\ and spectral resolution was better than 2 cm\textsup{-1}. Further Raman experiments and maps were done using a LabRamHR800 (JobinYvon Horiba) with 532\,nm excitation wavelength.
The chemical composition was monitored \emph{in situ} by X-ray photoelectron spectroscopy (XPS, h\(\nu\)= 1486\,eV).\emph{ Ex-situ} high resolution synchrotron radiation XPS measurements were performed in the SOLEIL synchrotron facility, Saint-Aubin, France, using photons with the energy of 350\,eV and 600\,eV.
Topography of the film was assessed by atomic force microscopy (AFM). Air AFM images were taken with Digital Instruments NanoScope III device. 
Local transport experiments were performed by means of a four-tip scanning tunneling microscope (4-tip STM) in combination with a high resolution scanning electron microscope (SEM).

\textbf{Theory.} Ab initio density functional theory (DFT) calculations for total energies, atomic structures, energy barriers, and molecular dynamics (in the range of picoseconds) have been performed using Quantum Espresso.\cite{quantespresso} Generalized Gradient Approximation (GGA) in the Perdew, Burke and Ernzerhof formulation\cite{PBE} was used for the exchange and correlation energy. Ultrasoft potentials were used to lower the energy cutoff down to 30\,Ry. The reciprocal space was sampled in two special points of the Brillouin zone of 4$\times$4 Ge(001) surface area. Activation energies were obtained by the Nudged Elastic Band algorithm\cite{NEB} and refined with the Climbing Image approach.\cite{NEB_CI} Ge slabs consisting of 8 Ge(001) layers, terminated on one side with H atoms, and separated with up to 3\,nm of vacuum were used; dipole correction was applied to decouple the slabs.

\textit{Conflict of interest}. The authors declare no competing financial interest.

\textit{Acknowledgement} Atomistic calculations have been done at the J\"ulich Supercomputing Centre, Germany, NIC project hfo06. The authors thank Ion Costina, Oksana Fursenko, David Kaiser, Wolfgang Mehr, Hans Thieme, Dominique Vignaud, Damian Walczyk, and Andre Wolff for technical support and discussions. Support from the European Commission through a STREP project (GRADE, No. 317839) is gratefully acknowledged. F.H. and J. M. acknowledge support by the European Research Council (ERC), grant number 259286. 

%
%
%
%
%

\vspace{0.4cm}

\textit{The bibliography section} is located after the Supporting Information section. %

\vspace{1.5cm}

\section*{\huge Supporting Information}
\vspace{0.7cm}

\textit{The following Supporting Information is available:} %
XPS measurements: comparison of C\,1s, Ge\,3d, and Si\,2p spectra obtained at various stages of C deposition on Ge(001) and Si(00), and detailed description of the SR-XPS study. 
Raman spectroscopy: analysis of Raman peaks, statistical data.
Electrical measurements: 4-tip STM setup, temperature dependence and Anderson localization, IV curves, SEM images.  
Ab initio calculations: expected growth modes, calculated energy barriers, kinetics of energy dissipation after C adsorption, comparison of C in the bulk of Ge and on the surface of Ge.
%

\subsection*{SI 1. X-ray Photoemission Spectroscopy}
 
\begin{figure}[th]
\mbox{}\hfill
\includegraphics[width=0.93\columnwidth,clip, trim=0mm 0mm 0mm 0mm]{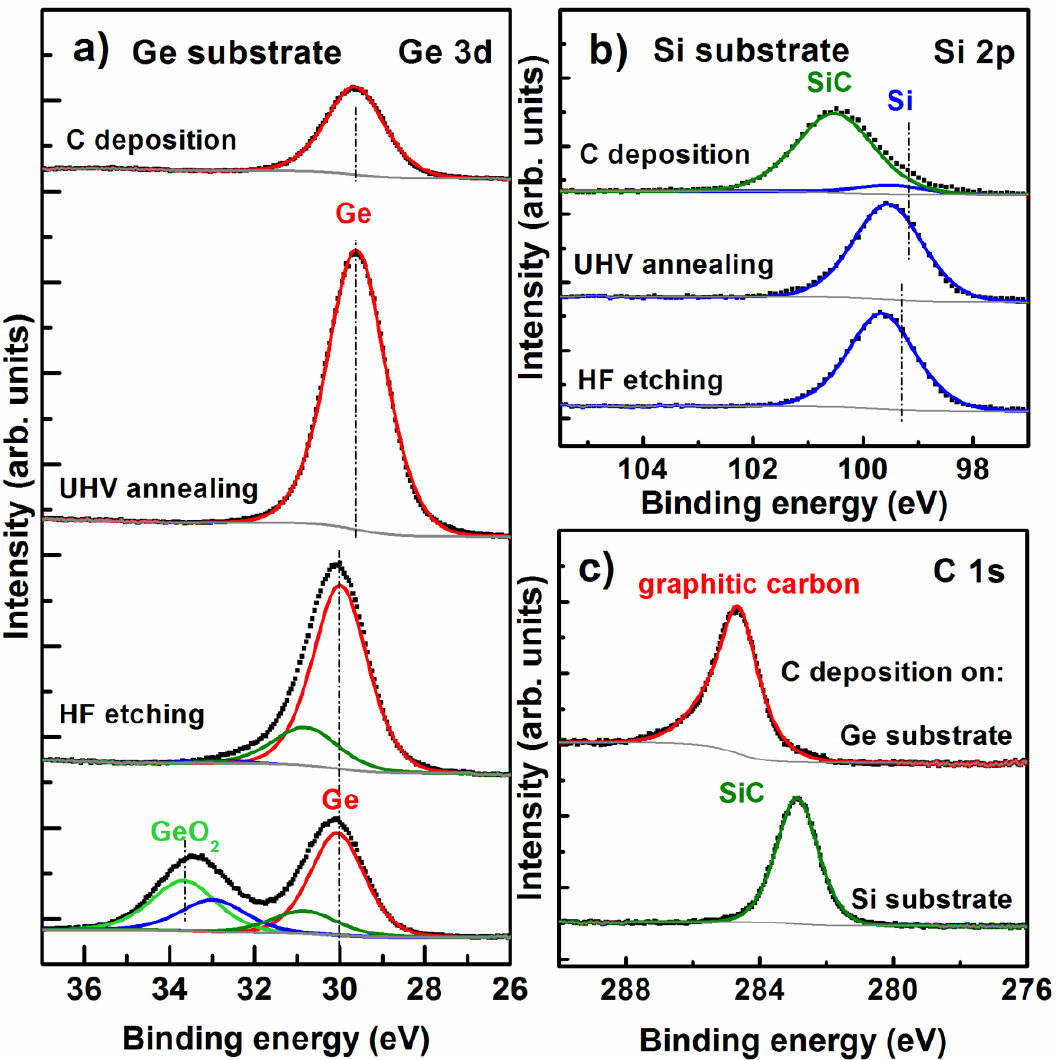}
\hfill\mbox{}
\caption{\label{SIfig:xps_new} In situ XPS investigation of C deposition on Si and Ge. (a) Ge 3d spectra for the Ge substrate at various stages. (b) Si 2p spectra for the Si substrate at various stages. (c) comparison of C 1s spectra for Ge and Si substrates after carbon deposition at $850^\circ$C. XPS chemical shifts are attributed to the formation of carbide on Si substrate and graphitic carbon on Ge substrate.}
\end{figure}

Chemical composition of the substrate surfaces and the deposited layers was investigated in-situ by x-ray photoelectron spectroscopy (XPS). Figure $1$ shows XPS measurements performed during preparation of the Ge and Si substrates and after deposition of few monolayers of carbon at $850^\circ$C. Native oxide layers in both cases is removed by HF dip. This standard cleaning procedure results in a clean oxygen-free Si surface (Fig.\,\ref{SIfig:xps_new}b), however, a residual signal from a substoichiometric oxide on germanium\cite{Molle2006,sun2006} is still detected  (Fig.\,\ref{SIfig:xps_new}a). This Ge suboxide is effectively removed by a short annealing at $750^\circ$C providing a clean Ge surface. Deposition of carbon on the Si surface results in a chemical shift of the main Si 2p photoemission line ($\Delta$E = 1 $\pm 0.1$ eV) which is attributed to the formation of SiC \cite{hackley,landoltSiC}. In contrast, the Ge 3d peak (29.6 eV) only decreases in intensity upon C deposition but no chemical shift is observed. In particular, no change in the position of the Ge 3d line proves that carbide formation does not take place which is in line with the Ge-C binary phase diagram \cite{CGePhasediag}. Figure 3c compares the C 1s spectra on both samples measured after C deposition. Based on the peak positions (282.8 eV and 284.7 eV for Si and Ge substrates, respectively) and shapes (symmetric on Si and asymmetric on Ge) we conclude that the C deposit on Si substrate is converted into silicon carbide, while on Ge it takes the form of graphitic carbon \cite{xiaolinLi}. 


SR-XPS C 1s core level spectra (cf. Fig.\,3b in the main part) were obtained by measurements performed in the SOLEIL synchrotron facility. The spectra were taken for graphene films growth on different substrates: Cu foil, SiC(000$\overline{1}$) (C face), SiC(0001) (Si face), and on Ge(001). The spectra were fitted accordingly with several peaks that describe various chemical carbon functionalities by taking into account the combined instrumental resolution of the experimental setup. The graphene component (green) has been fitted with a Doniach-Sunjic function which best reproduces the asymmetry on the higher binding energy side. The asymmetry on the higher binding energy side implies graphene structure and metallic conductivity. This can be measured by the singularity asymmetry factor $\alpha$, which is related to the delocalization of the valence states. The green (graphene) component  has an asymmetry factor a = 0.07. The spectra corresponding to SiC(0001) substrate consist of a SiC bulk component (in red), the graphene component (green) and two well-known interface contributions (blue). 

The information depth was about two monolayers. The important energy to take into account for determined the mean free path (MFP) of the photo-emitted electrons is the kinetic energy (KE) of the electrons. The photon energy used in the experiments was 350\,eV for all substrates in exception of graphene on C face of SiC, where 600\,eV photon energy was used. In the case of 350\,eV photon energy, the KE of the C\,1s core level is 61\,eV. For 600\,eV photon energy, the KE of the C1s is 326\,eV. For electrons of 60\,eV the MFP is close to 0.4-0.5\,nm (1-2\,monolayers) For electrons of 320\,eV the MFP is 0.6-0.7\,nm (near 2\,monolayers).

\textbf{Graphene on Cu:} graphene was grown by Chemical Vapor Deposition of graphene on copper foils purchased from Alfa Aesar (50 mm thickness, 99.9995\% purity). The details of the growth process are described elsewhere\cite{Wang2013}.

\textbf{Graphene on C-face of SiC:} graphene was grown on nominally on-axis SiC substrates with the aim of obtaining graphene samples with a thickness from 1 or 2 to about 10 monolayers. The substrate was production grade n-type 6H from SiCrystal and was cleaned using the standard RCA cleaning procedure before introduction into the sublimation furnace. One sample was prepared on an on-axis 4H substrate. High temperature sublimation with a buffer inert gas was used. The  temperature range was 1800-2000\degC, the pressure range was 500-850\,mbar, and the average growth time was 15\,min.

\textbf{Graphene on Si-face of SiC}: the sample was prepared by standard graphitization at IEMN, Universit\'{e}\ Lille~1. The graphitization of the SiC substrate at 1220\,degC\ for six minutes produced a high quality bilayer graphene on the Si-terminated SiC(0001) surface. A more detailed description of the growth is described elsewhere.\cite{XPS11}


\subsection*{SI 2. Raman Spectroscopy}

\begin{figure}[t]
\includegraphics[width=0.95\columnwidth,clip, trim=0mm 0mm 0mm 0mm]{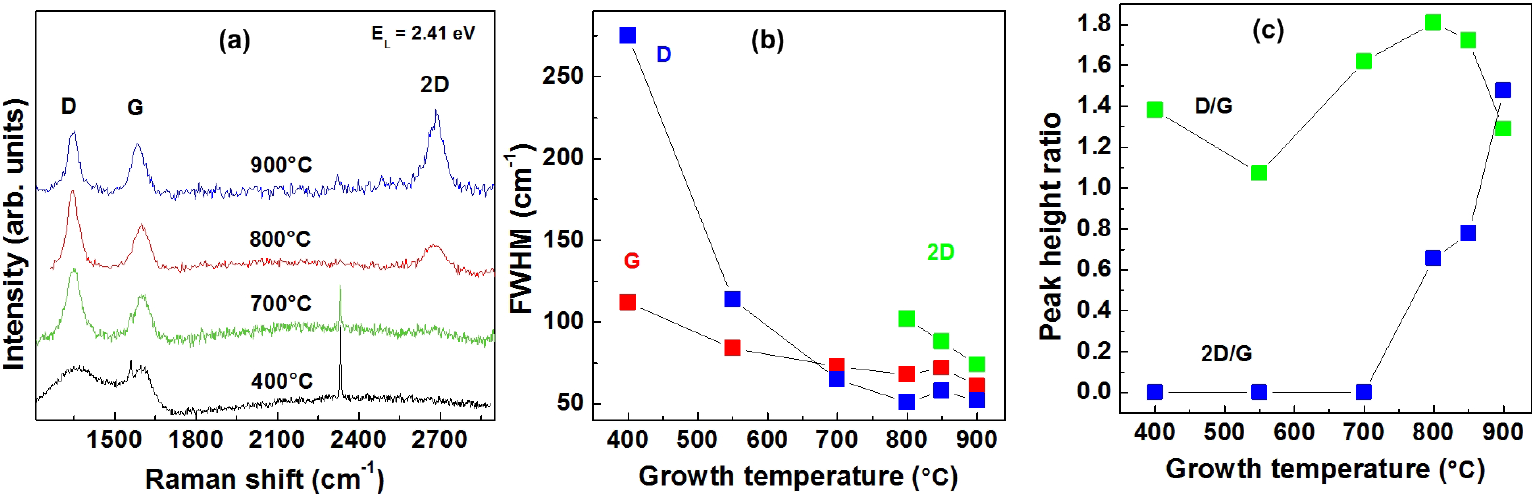}
\caption{\label{SIfig:raman} (a) Raman spectra from samples grown at at various substrate temperatures. (b) Peak width dependence on the growth temperature. (c) D/G and 2D/G peak height ratio dependence on the growth temperature.}
\end{figure}

Ex-situ Raman spectroscopy measurements on a series of samples grown at various temperatures and having nominal thickness of 5\,monolayers of carbon are summarized in Fig.\,\ref{SIfig:raman}. Figure\,\ref{SIfig:raman}a shows the Raman spectra. The sample prepared at $400$\degC\ exhibits a broad peak over the 1000-1700\,cm$^{-1}$ range, which is an indicative of amorphous C. Increasing the growth temperature to $700$\degC\ results in the transformation of this broad feature into two distinct peaks centered at around 1345 and 1600\,cm$^{-1}$ (D and G, respectively). Furthermore, a very weak local maximum can be recognized at 2679\,cm$^{-1}$. It  develops to a well defined peak (2D) as the growth temperature is raised to 800\degC. Presence of the characteristic D, G, and 2D bands proves the growth of sp$^2$-hybridized carbon on Ge substrates already in the $700-800$\degC\ temperature range. Narrowing of all peaks is observed when growth temperature is further increased (Fig.\,\ref{SIfig:raman}b), indicating the improving crystalline order. In addition, the 2D band gains intensity with respect to the D and G bands and at $900$\degC\ the 2D peak becomes dominant in the spectrum.

Due to the versatility of Raman spectroscopy for studying all kind of graphitic systems \cite{Reich2004}, we also performed Raman measurements with a high spectral resolution on single spots, as well as large-area Raman mappings on the MBE-grown graphene samples on germanium. For comparison, we exfoliated graphene from natural graphite on Ge substrates. The substrates used in the exfoliation process and MBE growth are the same, however,  the germanium for the exfoliation process was not heated in contrast to the germanium for the growth process. The typical Raman spectrum of graphene consists of the first-order $\Gamma$-point E$-{\rm 2g}$ phonon around 1580\,cm$^{-1}$ (G mode) and several double-resonant Raman modes, namely the D, \Dprime, and 2D mode around 1350\,cm$^{-1}$, 1620\,cm$^{-1}$, and 2670\,cm$^{-1}$, respectively (for the laser excitation energy of 2.33\,eV). These Raman peaks result from a second-order double-resonsance process and can be activated either by a defect (D and \Dprime\ mode) or a second phonon (2D mode). Thus, the intensities of the defect-related Raman modes can be used to investigate the structural and crystalline quality of the grown layers. The 2D mode stems from an intervalley double-resonant scattering process involving two TO phonons close to the K point \cite{thomsen2000}. Since the double-resonance process depends on both the electronic band structure and the phonon dispersion, the 2D-mode lineshape gives information about both properties \cite{ferrari2006}. 

\begin{figure}[th]
\includegraphics[width=1\columnwidth,clip, trim=0mm 0mm 0mm 0mm]{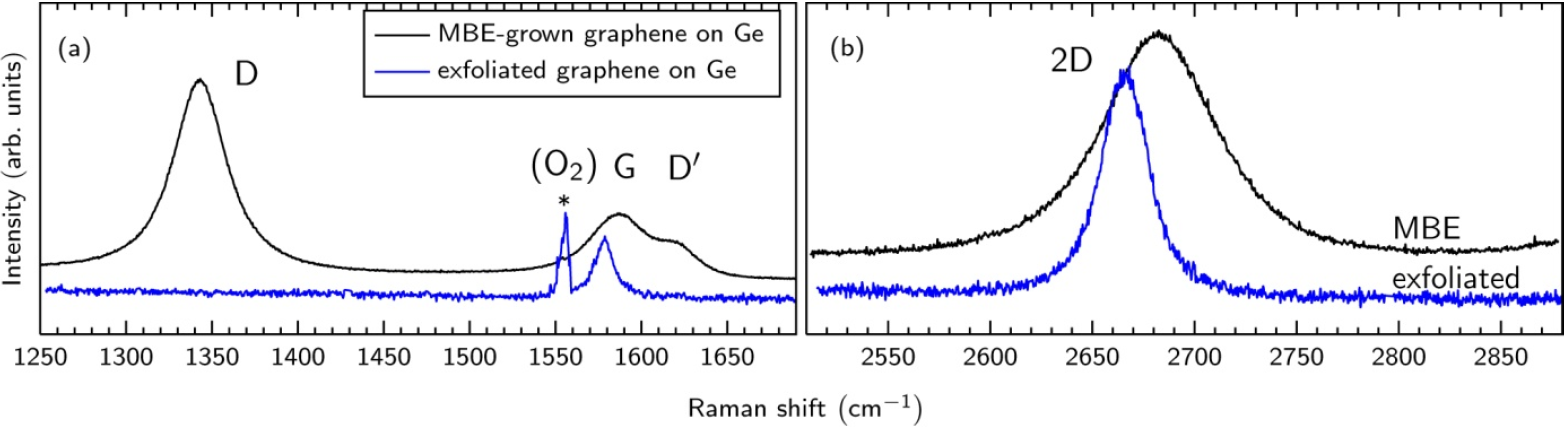}
\caption{\label{SIfig:raman_x} Raman spectra of the MBE-grown graphene 
 (black, upper spectra) and exfoliated single-layer graphene (blue, lower spectra) on germanium in the range of the D and G mode (a) and in the range of the 2D mode (b). Spectra are normalized to the same G and 2D mode intensity, respectively, and vertically offset for clarity. The asterisk marks the Raman peak of atmospheric oxygen (O$_2$) that appears due to long integration times.}
\end{figure}

\begin{figure}[th]
\includegraphics[width=1\columnwidth,clip, trim=0mm 0mm 0mm 0mm]{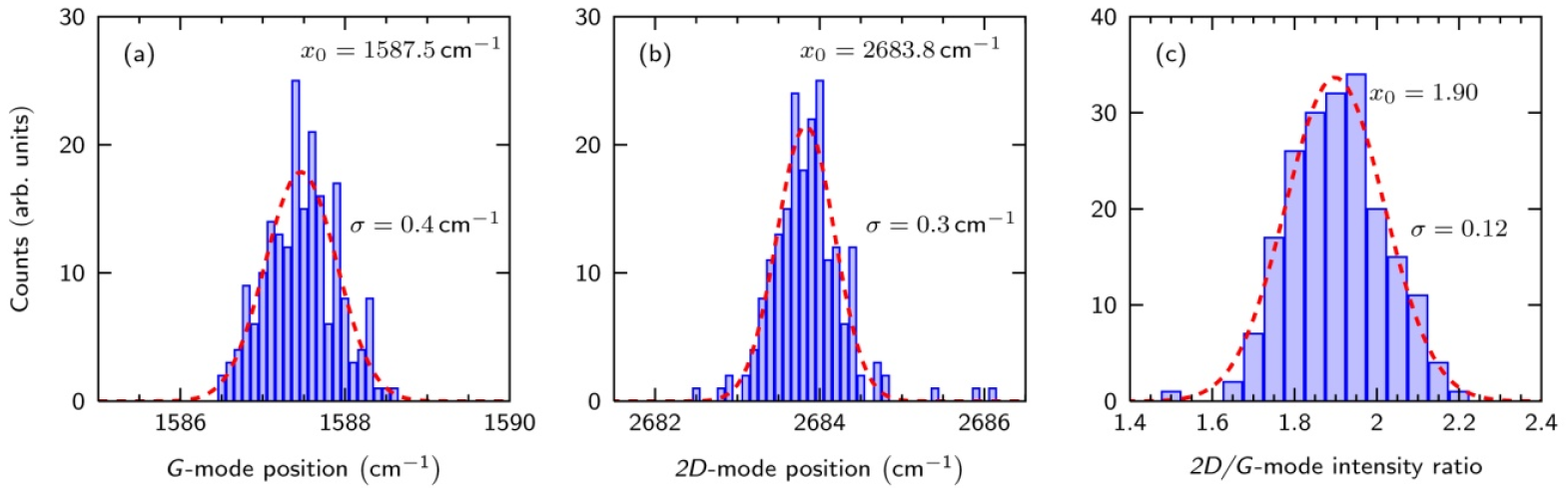}
\caption{\label{SIfig:raman_y}Statistical analysis of (a)~the G-mode, (b)~2D-mode position, and (c)~of the 2D/G-mode intensity ratio. The intensity of the G and 2D mode is evaluated from the peak areas. The data was fitted assuming a Gaussian distribution (red, dashed lines). 
Data collected from the area of 10$\times$30\,\um$^2$ with 200 spectra.
}
\end{figure}

Figure \ref{SIfig:raman_x} shows Raman spectra of a MBE-grown graphene sample together with spectra from exfoliated single-layer graphene on germanium. The spectrum of the grown sample shows all prominent Raman peaks of graphene. Since the D and 2D modes possess a breathing-like vibrational pattern of a single hexagonal carbon ring, the appearance of these Raman modes proves the growth of sp$^2$-hybridized carbon \cite{Tuinstra1970}. However, all Raman peaks are strongly broadened and up-shifted compared to exfoliated single-layer graphene on Ge. For instance, the G and 2D modes exhibit a FWHM of $\gamma_{\rm G}$  = 40.8$\pm$1.0\,cm$^{-1}$ and $\gamma_{\rm 2D}$ = 62.6$\pm$0.9\,cm$^{-1}$, respectively, and a peak position of $\nu_{\rm G}$ = 1587.5$\pm$0.4\,cm$^{-1}$ and $\nu_{\rm 2D}$ = 2683.8$\pm$0.3\,cm$^{-1}$ (see Fig.\,\ref{SIfig:raman_y}). In contrast, the mechanically exfoliated single-layer graphene on Ge exhibits values of $\gamma_{\rm G}$ = 1578\,cm$^{-1}$ and $\nu_{\rm G}$ = 13\,cm$^{-1}$ for the G mode and $\gamma_{\rm 2D}$= 2666\,cm$^{-1}$ and $\nu_{\rm 2D}$ = 26\,cm$^{-1}$ for the 2D mode. The positions and FWHMs of the Raman modes for exfoliated single-layer graphene on germanium resemble the typical values found for free-standing (undoped, unstrained) single-layer graphene \cite{berciaud2009}. 
The up-shift and broadening of the Raman modes in the MBE-grown sample indicate the presence of nanocrystalline graphene \cite{Ferrari2000}. In general, Raman spectroscopy is used to evaluate the doping level or the amount of strain in graphene by  the G- and 2D-mode position and their FWHM. In the present case this is rather difficult, since the influence of the defects superimposes possible effects from strain or doping. Instead, the up-shift of the Raman modes is most likely related to the nanocrystallinity of the graphene (see below). 

The 2D mode of the MBE-grown graphene is symmetric with a broad single-Lorentzian shape. However, the symmetric line shape does not indicate single-layer graphene in the grown samples but rather few-layer graphene. It was shown for CVD-grown graphene \cite{Reina2009} and recently for MBE-grown graphene on sapphire \cite{Oliveira2013} that in such samples the 2D-mode line shape remains symmetric up to approximately three layers but up-shifts and broadens. Since the 2D-mode line shape reflects to a certain extent the evolution of the electronic band structure of few-layer graphene around the K points, this single-Lorentzian shape in thicker graphene samples indicates that the grown graphene layers do not have a defined stacking order and thus are decoupled \cite{Poncharal2008}. Since the grown graphene layers are decoupled, we cannot reliably determine the exact number of layers just from the 2D-mode line shape or by the appearance of other layer-number dependent Raman modes \cite{ferrari2006,Herziger2012,Lui2013}. However, we can conclude that the MBE-grown graphene samples are not single-layer graphene. Furthermore, we can exclude the growth of carbon nanotubes, since no radial-breathing modes (Raman spectra not shown) nor a splitting of the G mode in G$^-$ and G$^+$ were observed (see Fig.\,\ref{SIfig:raman_x})\cite{ReichBook2004}.

To evaluate the quality of the grown graphene samples, we performed Raman mappings covering an area of approximately 10$\times$30\,\um$^2$ with a total of 200 spectra. The statistical analysis of the peak position of the G and 2D mode is shown in Fig.\ref{SIfig:raman_y}a and \ref{SIfig:raman_y}b. The data was fitted assuming a Gaussian distribution. The low standard deviation ($\sigma$=0.4\,cm$^{-1}$) of the experimental data from the peak positions $\gamma_{\rm G}$ = 1587.5\,cm$^{-1}$ and $\gamma_{\rm 2D}$ = 2683.8\,cm$^{-1}$ verifies the very high homogeneity of the grown graphene layers. Figure \ref{SIfig:raman_y}c shows an analysis of the 2D/G-mode intensity (integrated area) ratio in the investigated region, which has an average value of 1.90$\pm$0.12. This ratio is considerably lower than the values found for single-layer graphene, which exhibits typically a ratio of approximately four at excitation wavelengths of 532\,nm and using SiO$_2$ substrates \cite{ferrari2006}. Again, this indicates that the MBE-grown graphene on germanium is not a single layer. The D/G-mode intensity ratio in the investigated area was determined to 2.19$\pm$0.07. It is well known that the D/G-mode ratio is indicative for the grain size $L_a$ of nanocrystalline graphene and graphite \cite{Tuinstra1970,Cancado2007,Lucchese2010}, where $L_a$ is defined as the average crystal diameter and given by

\begin{equation}
	L_a({\rm nm}) = (2.4\times 10^{10})\lambda^4_{\rm Laser} (\frac{I_{\rm D}}{I_{\rm G}})^{-1}
\end{equation}

\noindent Using the D/G-mode intensity ratio from our samples, the laser wavelength of 532\,nm used in our experiments, and applying the above equation, we deduce an average crystalline grain size of  $L_a$ = 8.8$\pm$0.3\,nm.

In summary, the Raman analysis proves the growth of sp$^2$-hybridized carbon. The MBE samples consist of decoupled few-layer graphene. From the intensity ratio of the D and G mode we deduced a grain size of the nanocrystalline graphene of approximately 10\,nm.

\subsection*{SI 3. Electrical measurements}

\begin{figure}[th]
\mbox{}\hfill
\includegraphics[width=0.8\columnwidth,clip, trim=0mm 0mm 0mm 0mm]{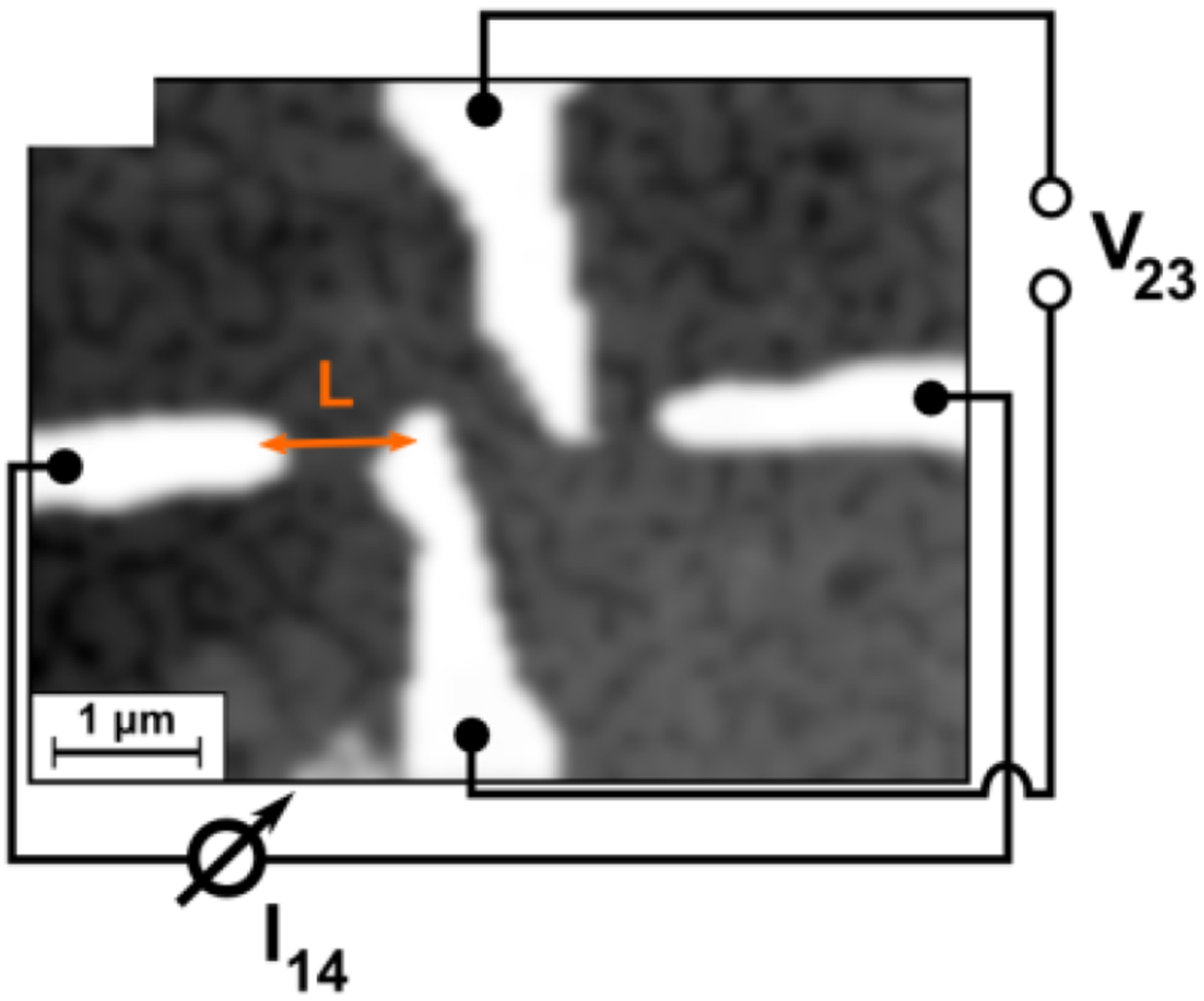}
\hfill\mbox{}
\caption{\label{SIfig:4STM} SEM image of the 4-point probe configuration.}
\end{figure}

Local transport experiments were performed by means of a four-tip scanning tunneling microscope (4-tip STM, Fig\,\ref{SIfig:4STM}) in combination with a high resolution scanning electron microscope (SEM). The SEM gives a quick access to the local morphology of the samples and allows to choose a specific area for further transport experiments. After the first SEM analysis, all four tips of the 4-tip STM are brought into tunneling contact by a feedback control loop and are then navigated to the desired positions. Operating in tunneling mode ensures that the tips do not touch the surface while they are moved over the sample. Once the preselected area for transport experiments is reached, the feedback is turned off and all tips are lowered one by one until they make direct contact with the sample surface. During this process, contact resistances between the individual tips are checked simultaneously to ensure that all tips are in contact with the sample. All IV curves are recorded in a 4-point probe geometry. Hence, a current is driven through the outer probes and the voltage drop between the inner probes is recorded. This allows us to measure the resistance of the sample without parasitic influences of contact resistances.

\begin{figure}[th]
\mbox{}\hfill
\includegraphics[width=0.85\columnwidth,clip, trim=0mm 0mm 0mm 0mm]{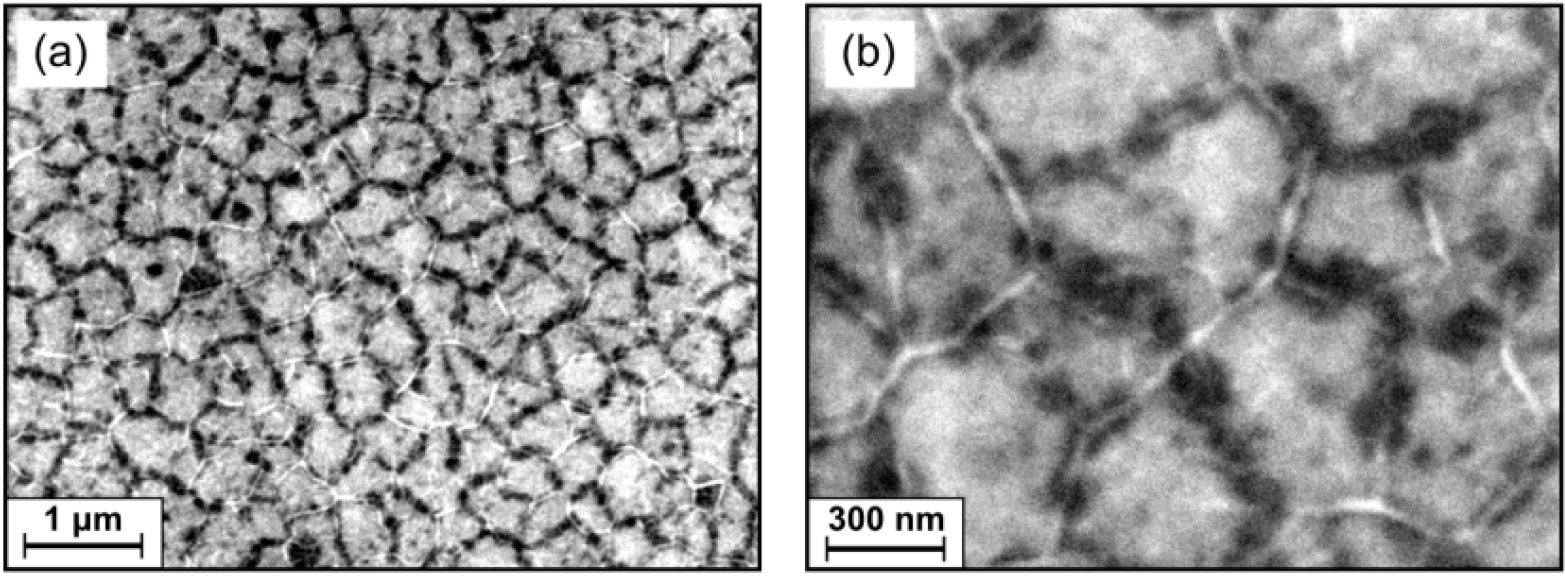}
\hfill\mbox{}
\caption{\label{SIfig:SEM} SEM images of sample was grown at the substrate temperature of 930\degC. The nominal coverage is 12 monolayers of carbon.
}
\end{figure}

Four samples with different preparation conditions (growth time from 200\,s to 1000\,s, growth temperatures of 900\degC\ and 930\degC) were investigated. Two SEM pictures on different scalings (Fig.\,\ref{SIfig:SEM}a-b) of the sample prepared at 900\degC\ with a growth time of 500\,s are shown exemplarily. The local morphology seen in the SEM images is consistent with AFM studies for all investigated samples.

\begin{figure}[b]
\mbox{}\hfill
\includegraphics[width=0.88\columnwidth,clip, trim=0mm 0mm 0mm 0mm]{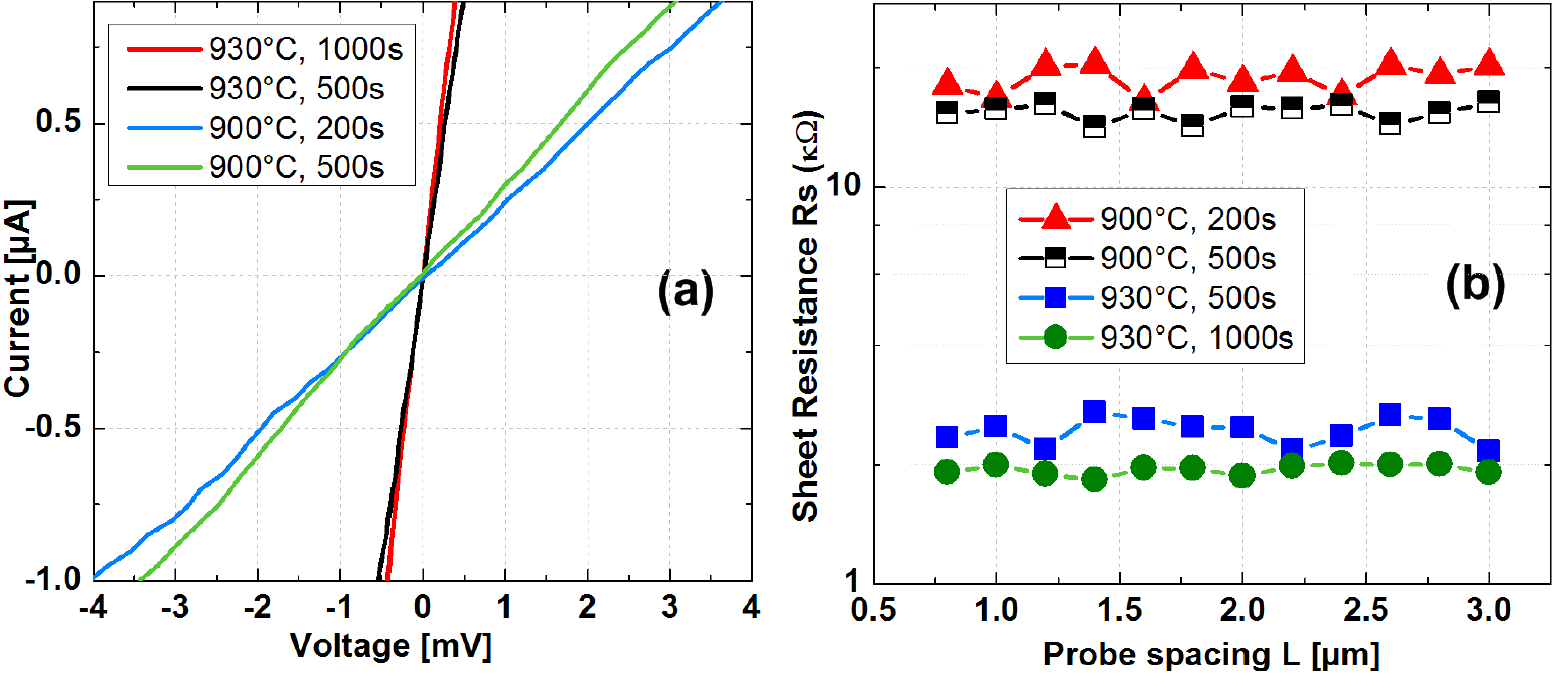}
\hfill\mbox{}
\caption{\label{SIfig:Electrical_1} Room temperature measurements. (a) IV curves, $L = 2$\,um (cf. Fig.\,\ref{SIfig:4STM}) on various samples. (b) Sheet resistance as a function of probe spacing.}
\end{figure}

Local transport experiments were realized with tip spacings ranging from 500\,nm to 3\,\um. All transport data were recorded in an equidistant linear arrangement of the probes. A typical arrangement of the tips with the electrical wiring shown schematically can be found in Fig.\,\ref{SIfig:4STM}. The recorded IV curves for a fixed probe spacing of 2\,um are shown in Fig.\,\ref{SIfig:Electrical_1}a for all four investigated samples. Most striking is the linearity of the data indicating clearly a metallic behavior of the surface. Already visible is also a strong difference between the samples processed at 930\degC\ and those processed at 900\degC, which show a much higher resistance. To get more information about the nature of the electronic transport, probe spacing dependent measurements were carried out (Fig.\,\ref{SIfig:Electrical_1}b). All investigated samples exhibit a probe spacing independent resistance which is a typical property of two dimensional transport \cite{Schroder1998}. No contributions of the bulk Ge were seen in any of the transport experiments. A possible influence of the Ge surface state will be discussed below together with the temperature dependence. The probe spacing independence of resistance allows us to calculate the two dimensional resistivity (sheet resistance $R_{\rm S}$) of the graphene crystallites in a very simple way with the expression \cite{Schroder1998}

\begin{equation}
	R_{\rm S} = \frac{\pi}{\ln(2)}  \frac{ U_{23}}{I_{14}}.
\end{equation} 

\noindent The sheet resistances calculated with this formula are plotted in fig. 1(e). The samples processed at the surface melting point at 930\degC\ exhibit sheet resistances about 2\,k$\Omega/\Box$ which are 10 times smaller than those measured on the samples prepared below the surface melting point. This reflects the better electronic quality of the graphene nanocrystallites prepared at the surface melting point. However, no significant difference between samples processed at the same temperature but with different growth times could be found. This might be an indication that either only a fixed number of graphene layers participates in electronic transport, or that the additional amount of carbon is fully implemented into the carbon islands rather than into the graphene nanocrystallites in between.


To explore the influence of the strong inhomogeneity of the sample on the local transport properties, temperature dependent measurements were carried out. For this purpose, the sheet resistance of the samples which showed the lowest resistivity (those processed at the surface melting point) were measured within a temperature range from room temperature down to 36 K (obtained by cooling with liquid He). The results are shown in Fig.\,6a in the main article. 
With decreasing sample temperature the sheet resistance is exponentially increasing from 2\,k$\Omega/\Box$ up to 20\,k$\Omega/\Box$. This behavior is well known for disordered systems and especially also for disordered graphene and can be understood in terms of Anderson localization \cite{Anderson1958,Peters2012,Yan2010,Cheah2013}. To verify this assumption we fit the data according to the model of variable range hopping (VRH) which is the basic transport mechanism in Anderson localized systems \cite{Mott1971}. For VRH the resistance is given by 

\begin{equation}
	\ln(R) \sim (\frac{T}{T_0})^{d+1},
\end{equation}
 
where $d$ denotes the dimension. From the inset in Fig.\,6a (main article) 
 it is already obvious that the data are best described with $d$ = 2, which is consistent with the observation of two-dimensional transport behavior. The full fit is also shown in Fig.\,6a (main article) 
 and describes the data very well on the complete temperature range. A possible contribution of the Ge surface state \cite{Kevan1985} is also excluded by the temperature dependent transport data. The measured IV curves maintain their metallic behavior even at the low temperature. On the contrary, the Ge(001) surface was reported to be either only metallic at temperatures above 200\,K and semiconducting below 200\,K \cite{Eriksson2008} or to be semiconducting even at room temperature \cite{Jeon2006}. In all cases, the temperature dependence of the surface state does not match our transport data, hence, a parasitic contribution of the Ge surface state can be excluded. Therefore we conclude that we probe only the electronic properties of the graphene and not of the underlying Ge substrate.

In summary, the local transport experiments have shown that the graphene nanocrystallites behave as a two-dimensional conductor. No contribution of the underlying substrate could be detected. The temperature dependent measurements reflect the inhomogeneity of the investigated samples showing strong localization effects. The data agree very well with the theoretical model of variable range hopping in two dimensions.

\subsection*{SI 4. Ab initio calculations}

Results of ab initio calculations for the behavior of C atoms on Ge(001) p(2$\times$2) surface point to three mechanisms as potentially responsible for the observed surface roughness. They are sketched in the next paragraph and discussed in more detail later on.

\begin{figure}[b]
\includegraphics[width=1\columnwidth,clip, trim=0mm 0mm 0mm 0mm]{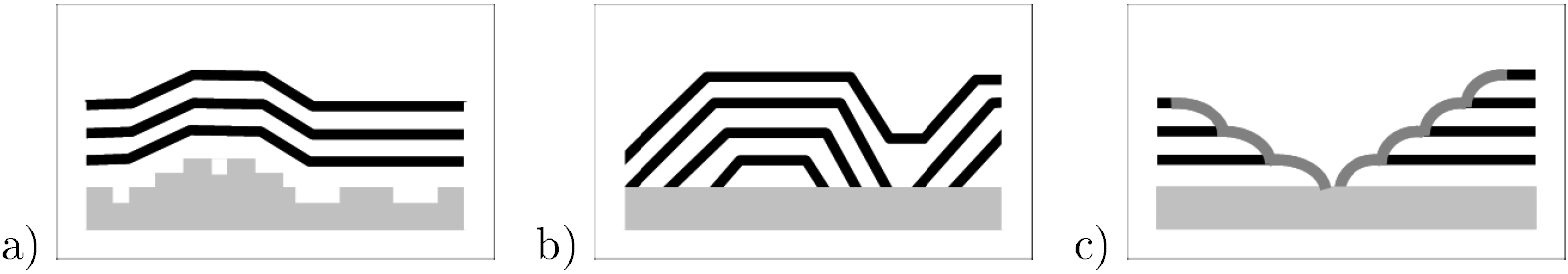}
\caption{\label{fig:Roughening_mechanisms} Expected result of graphene growth dominated by each of the three roughening mechanisms discussed in the text. (a) Germanium ejection might lead to graphene covering a roughened Ge substrate. (b) Chemical interaction between graphene edge and substrate dimers might lead to formation of multilayer graphene islands with bonded edges. (c) Migration of ejected Ge to graphene edges might lead to formation of Ge-rich facets terminating the multi-layer graphene islands. }
\end{figure}

The first plausible reason for the roughness is ejection of Ge atoms from surface dimers by C atoms. This implies that during the first stage of deposition many carbon atoms are immobilized in surface substitutional sites, while the Ge atoms ejected from these sites agglomerate into islands, possibly with some of the deposited carbon atoms. During the second stage of growth, graphene nucleates and grows on top of such a pre-roughened surface (Fig.~\ref{fig:Roughening_mechanisms}a). On the other hand, when one of the other two mechanisms dominates, the substrate surface retains its flatness. With the mechanism number two, the graphene growth mode is strongly influenced by the formation of chemical bonds between the edge of the first layer of graphene flakes and the topmost atoms of germanium (Fig.~\ref{fig:Roughening_mechanisms}b). If this is the case, the aspect ratio of multilayer graphene islands is likely to depend on the crystallographic direction, along which the carbon atoms from the first graphene layer formed a stable chemical connection to the substrate. Finally, the mechanism number three works by combining the ejection of Ge and Ge-graphene interaction. This combination may lead to Ge contamination of graphene edges and to formation of Ge-rich facets on edges of multilayer graphene (Fig.~\ref{fig:Roughening_mechanisms}c).

\begin{figure}[th]
\mbox{}\vspace{5mm}
\includegraphics[width=1.0\columnwidth,clip, trim=0mm 0mm 0mm 0mm]{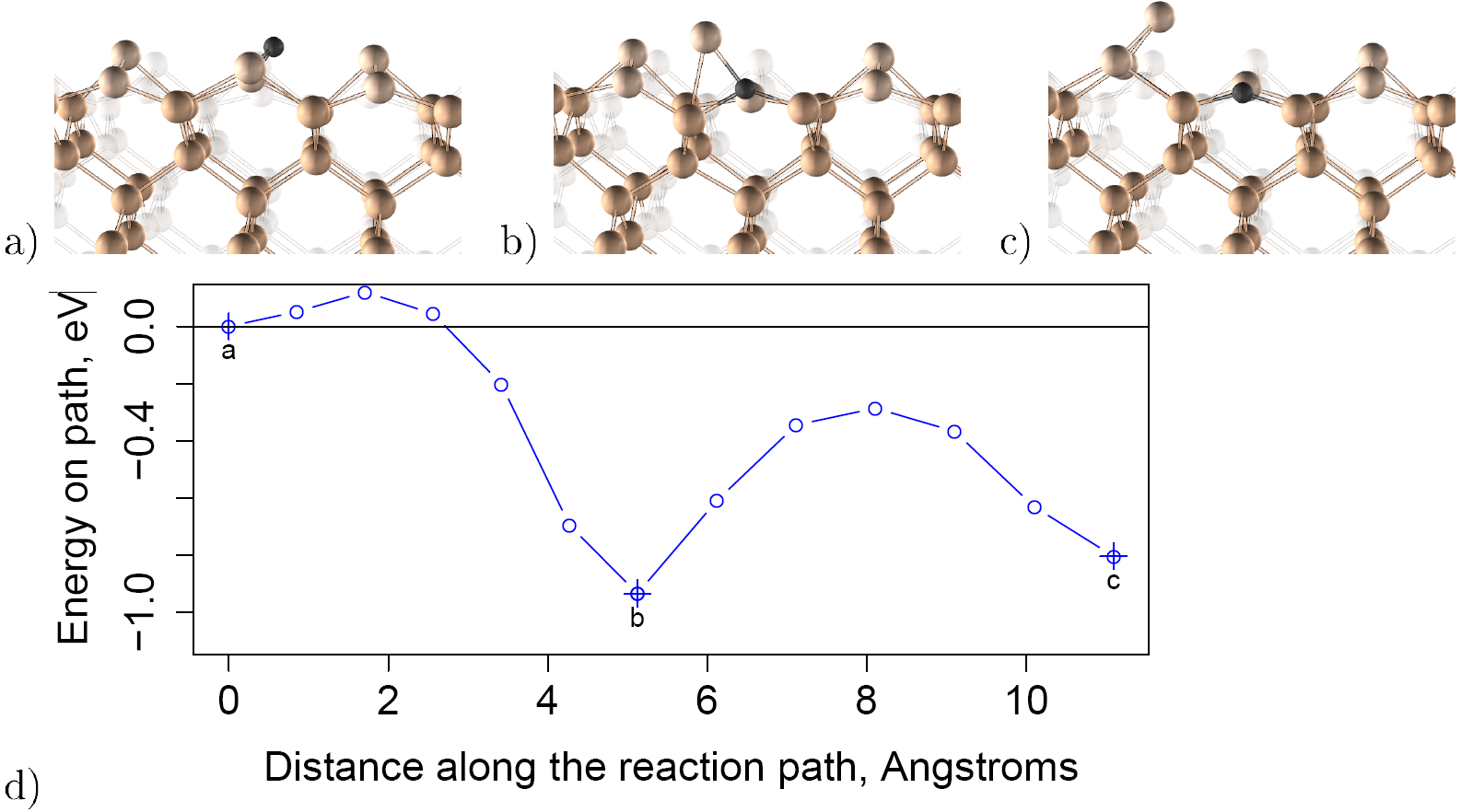}
\caption{\label{fig:GeEjectionFromDimer} Adsorption of C on Ge(001) and Ge ejection. (a-c)~Atomic configurations: (a)~a C atom (black) is adsorbed in a Ge dimer bond, (b)~it moves into the nearest interstitial site, elevating one of the Ge dimer atoms, and (c)~the Ge atom is ejected to the top of a neighboring dimer; the Ge-Ge dimer transforms into a C-Ge dimer. (d)~Energy barriers; the labels correspond to the atomic geometries above. Adsorption of atom from vacuum to the geometry "a" proceeds with no barrier, the energy gain is about\,4.6 eV. \vspace{3mm}}
\end{figure}

Ge ejection (Fig. \ref{fig:GeEjectionFromDimer}) is expected to take place because the energy barrier for this process is only about 0.65\,eV. This is low enough to allow for ejection of Ge at any practicable growth temperature. For example, assuming the attempt frequency of 10$^{13}$s$^{-1}$, the expected time needed to overcome this barrier is less than 0.1\,ns at 850$^\circ$C and less than 5\,ns at 450$^\circ$C. Furthermore, the barrier height is much smaller than the 4.6\,eV released by C adsorption from vacuum to the top of the surface, meaning that from the energy point of view the Ge ejection process may take place athermally or may be athermally assisted. Namely, the energy needed to overcome the barrier, or part of this energy, may be taken from the adsorption energy and not from thermal vibrations, meaning that the ejection probability does not depend on substrate temperature or that the effective barrier showing up in the Arrhenius plot is smaller than the physical barrier.

Ge monomers are highly mobile. Thus, if the ejection process is indeed efficient, it is plausible that they agglomerate into islands. Such islands will grow provided that the monomer diffusion length is significantly smaller than on the clean surface. Otherwise, the surface will evolve as during molecular beam epitaxy of Ge on Ge(001): few islands will nucleate and flat Ge will re-grow by step flow. One may speculate that the diffusion length of Ge monomers is reduced by interaction between the monomers and C atoms.

\begin{figure}[th]
\includegraphics[width=0.95\columnwidth,clip, trim=0mm 0mm 0mm 0mm]{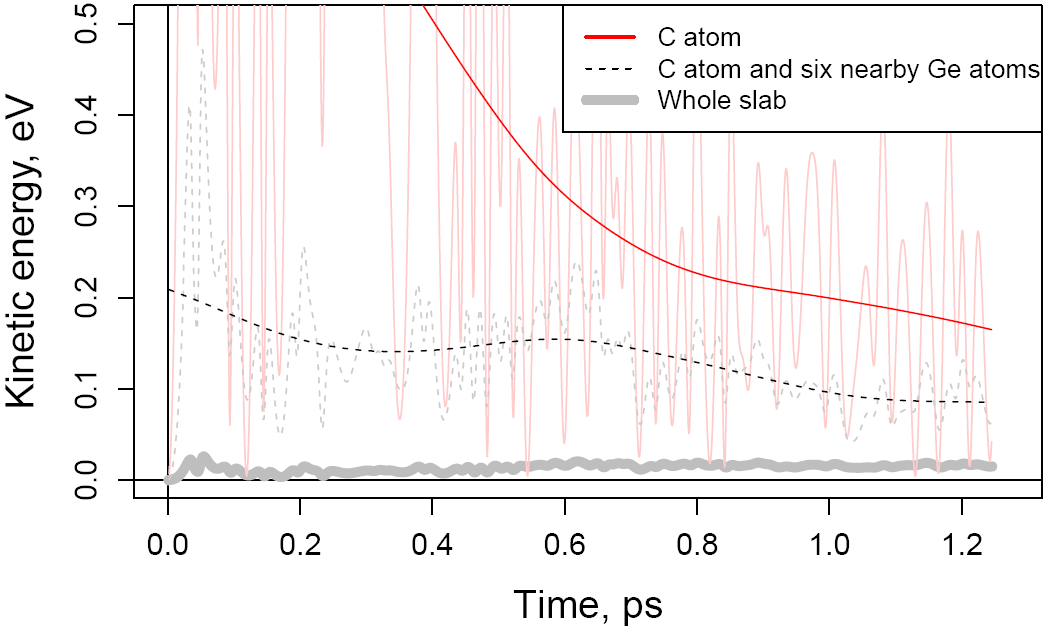}
\caption{\label{fig:MolecularDynamics} Kinetic energy of a carbon atom and average kinetic energy of seven Ge atoms in its nearest neighborhood, as a function of time elapsed from the moment of its first collision with the surface. Momentary and time averaged values are shown. At $t=0$\,ps the slab was at $T=0$\,K; the thick line shows the evolution of the average kinetic energy of slab atoms. }
\end{figure}

In spite of high adsorption energy of C from vacuum to Ge(001), athermal emission of Ge seems to be a rare event. Figure\,\ref{fig:MolecularDynamics} illustrates a typical evolution of the kinetic energy of a C atom during the first picosecond after adsorption. The average kinetic energy of nearby Ge atoms is plotted as well. We performed about a dozen simulation runs; in all cases the C atom did not diffuse away from the impact site further than a few {\AA}ngstroms before it finally occupied the sub-dimer interstitial site and thermalized to the energy well below the kinetic barrier for Ge ejection. Within the first picosecond, the impact energy was distributed among the neighboring Ge atoms, but the kinetic energy acquired by a single Ge atom was always lower than the kinetic barrier and, within a fraction of a picosecond, it dropped below 0.2-0.1\,eV. One can associate this behavior with large mass ratio between light C and heavy Ge atoms. The surface processes caused by C adsorption on Ge(001) may thus be to good accuracy described as happening at thermal equilibrium.  

The ejection barrier is low as long as a C atom occupies the sub-dimer interstitial site (Fig~\,\ref{fig:GeEjectionFromDimer}b). The efficiency of the carbon-assisted ejection of Ge depends therefore (a)~on the probability that the reaction of C with Ge proceeds through the path containing the sub-dimer interstitial, and (b)~on the total time spent by the C atom in this interstitial site. Indeed, C$_{\rm i}$ may in principle drift away from the sub-dimer interstitial geometry (Fig.\,\ref{fig:GeEjectionFromDimer}b) before its Ge neighbor had a chance to escape to the geometry shown in Fig. \ref{fig:GeEjectionFromDimer}c, or it may diffuse deeper into the bulk of germanium, omitting the sub-dimer interstitial site altogether. 


Figure 9b of the main part 
summarizes the energies of carbon adsorbed at and under Ge(001) p(2$\times$2) surface. It is apparent that C dissolved in Ge is unstable with respect to phase separation into graphene and clean Ge. The energy difference of about 2\,eV between C in graphene and substitutional C in bulk Ge is compatible with very low solubility of C in Ge (up to 2.5$\cdot$10$^{14}$\,cm$^{-3}$ in CZ-grown crystals,\citep{CinGsolubilityHaller1982} which corresponds to 2.0 eV at Ge melting temperature of 937$^\circ$). %
It is also clear that subsurface sites are generally more favorable for carbon than sites in the bulk; this effect is much stronger for interstitial (C$_{\rm i}$) than for substitutional (C$_{\rm Ge}$) geometries. The presence of numerous energetically nonequivalent subsurface sites of the same class (interstitial or substitutional) is due to the strain field caused by dimerization of the surface. 

Closer examination of Fig.\,9b (from the main part) 
reveals that the subsurface interstitial site with the lowest energy is not the one from which Ge ejection may be initiated (label "B", cf. the atomic structure in Fig.\,\ref{fig:GeEjectionFromDimer}b), but one that is burried deeper under the surface (label "X", (001)-split interstitial, i.e. an atom sharing a lattice site with a host atom). The energy barrier from "B" to "X" is 0.9\,eV, that is, nearly the same as the barrier for Ge ejection from "B". This means that, entropy factors ignored, it is expected that in about 90\% of cases Ge atom is thermally ejected before the C atom leaves the sub-dimer site "B". 

The carbon atom that has reached the burried "X" site does not eject Ge so easily. In order to do so, it would either have to return to the "B" site, or kick out and replace a neighboring Ge atom. The former process is associated with a kinetic barrier of 1.7\,eV, meaning one successful return attempt per $\tau_{\rm xb}$ = 4\,$\mu$s at 850$^\circ$C and one per $\tau_{\rm xb}$ = 70\,ms at 450$^\circ$C; hence, this process takes much longer than about nanosecond needed from Ge ejection from "B". Going from "X" to the most favorable subsurface substitutional site ("S" in Fig.\,9b in the main part) 
is even more time consuming. The corresponding barrier is nearly 3.0\,eV, which translates into one successful substitution attempt per second at 850$^\circ$ and one per year at 450$^\circ$C. 

Deposited C atoms are thus expected to accumulate in close vicinity to the Ge(001) surface, partially in form of C-Ge dimers substituting Ge-Ge dimers (as in the structure "c", shown in Fig.\,\ref{fig:GeEjectionFromDimer}c), and partially in form of interstitials (structure "X"). The "X" interstitial is highly mobile and it may also convert to other states, with various probabilities. It may substitute a fourfold-coordinated Ge atom, but this happens very rarely and is irrelevant for our discussion. Migration of the interstitial into deeper regions of Ge seems to be a rare event as well: in spite of low migration barrier of C$_{\rm i}$ in bulk Ge (our calculations indicate the barrier of 0.9 eV, which is practically the same as the barrier for C$_{\rm i}$ migration in Si),\cite{Pedersen1999} the energy of bulk C$_{\rm i}$ is substantially (by nearly 2.4\,eV) higher than the energy of subsurface C$_{\rm i}$. However, the conversion of "X" to a C-Ge dimer (with Ge ejection from the surface dimer) is of importance, because it can happen within times several orders of magnitude shorter than the time $\tau_{\rm CC}$ that elapses between another C atom from the beam hits within the intermediate neighborhood. The latter time is namely comparable to the reverse deposition rate measured in monolayers per time: to seconds or minutes, which is orders of magnitude longer than the life time  $\tau_{\rm x}$ of the "X" state.

The mobile C$_{\rm i}$ spends therefore most of its life at one of the "X" sites. It ceases however to exist within $\tau_{\rm x}$ of the same length scale as the characteristic time $\tau_{\rm xb}$ of the conversion from "X" to "B": microseconds at 850\degC\ and milliseconds at 450\degC. This is because the "B" state decays, with about 90\% probability, to a Ge surface monomer and a C-Ge surface dimer ((Fig.\ref{fig:GeEjectionFromDimer}c)). The highly mobile monomer diffuses then away. The interstitial at "X" might also end up by reacting with another C$_{\rm i}$, but since the time $\tau_{\rm CC}$ is much longer than $\tau_{\rm x}$, all interstitials are expected to be already in the C-Ge state before encounter with another C takes place. 

The C-Ge surface dimer is likely to live until another Ge monomer kicks C from it; the corresponding barrier amounts to about 0.5\,eV, it is therefore very low. For this reason we also suppose that overgrowth of the C-Ge dimer by a progressing surface step is quite improbable; what happens is rather that C is kicked out and either goes interstitial (to an "X" site") or ejects another Ge atom from another Ge-Ge surface dimer.

It follows that when two C atoms meet on Ge(001) surface, one of them has already substituted Ge in a surface dimer and is in the C-Ge dimer state. It seems plausible that this second C atom makes a bond to the C atom in the C-Ge dimer and that the whole object remains in its place until one more atom arrives. This can be either C or Ge. If C arrives, a C$_3$ cluster consisting of C in C-Ge dimer and of C$_2$ attached to it is expected to form. If Ge arrives, we suppose that a kick-in process takes place: Ge substitutes the C atom in the C-Ge dimer and a surface C-C dimer appears on top of the otherwise (locally) perfect Ge(001). The substitution of C in the C-Ge dimer a Ge monomer is however plausible only when the C cluster is small enough so that it does not block the access to the C-Ge dimer site. If the Ge monomer cannot approach the C-Ge dimer, it sticks to the C atoms at the edge of the growing cluster. The same is expected to happen when a Ge monomer encounters a C cluster sitting on top of the perfect surface: it attaches itself to the cluster edge. 

This analysis indicates that Ge contamination (Fig.\,\ref{fig:Roughening_mechanisms}c) is a more realistic reason for film roughness than substrate roughening (Fig.\,\ref{fig:Roughening_mechanisms}a), albeit at sub-monolayer C coverages the amount of ejected Ge atoms is comparable to the amount of deposited carbon. It also shows that the role played by Ge ejection is ambivalent. On the one hand, Ge monomers may contaminate the C clusters, possibly hindering the growth of crystalline graphene at lower temperatures, and possibly contributing to the formation of facets on the edges of growing graphitic islands. On the other hand, Ge monomers may heal the surface from the defects at which the graphene seeds are strongly attached to the substrate. Which one of the two opposite actions dominates, it may depend on the growth conditions: on the substrate temperature and on the deposition rate.

\begin{figure}[th]
\includegraphics[width=1\columnwidth,clip, trim=0mm 0mm 0mm 0mm]{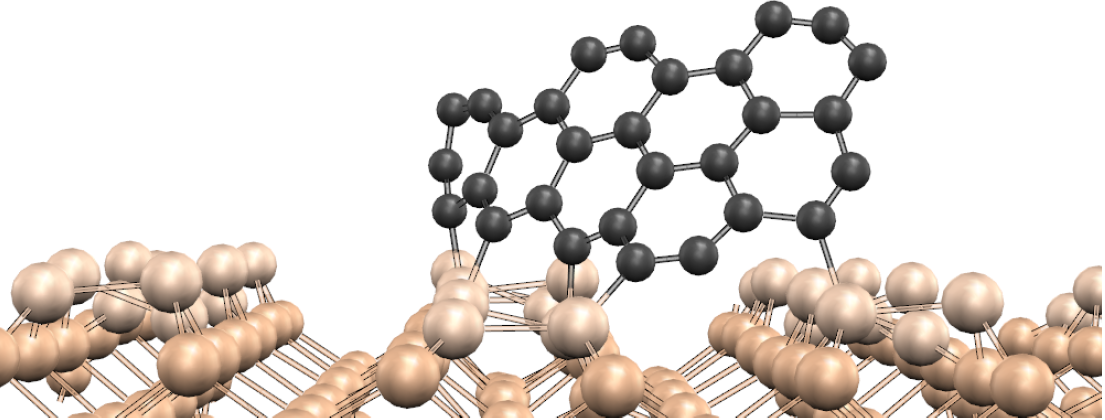}
\caption{\label{fig:C29} Formation of bonds between a small (C$_{29}$ graphene molecule on Ge(001)-2$\times$2.}
\end{figure}

So far we did not account for the possibility depicted in Fig.\,\ref{fig:Roughening_mechanisms}b: that graphene edge is bonded to the (otherwise perfect) substrate surface and that this effect limits the grain size. But when a small graphene molecule is placed on the Ge(001) surface, its edge atoms tends to make bonds with the substrate. Figure \ref{fig:C29} illustrates this for the case of a molecule consisting of 29 carbon atoms. The molecule attached itself to the substrate with five edge atoms: four C atoms in a sequence became bonded to Ge atoms from the same dimer row, and one more distant C atom became bonded to a Ge atom from a neighboring dimer row. The bond dissociation energy was in this case 1.5\,eV per C-Ge bond. Due to directional character of the bonds, the molecule experienced a force lifting it towards the surface normal. The effect is substantial and should have noticeable influence on the growth mode. Even though in its ground state a bigger molecule prefers to be parallel to the surface so that the number of C-Ge bonds at its edge is possibly high, many tiny C clusters will initially stand up, because there are many competing paths by which carbon clusters are built and grow, and some of these paths are likely to end up with a cluster that is bonded to the substrate only on one side. This non-parallel orientation is sustained by further C atoms. Namely, they reach the cluster in the form of "X" interstitials, i.e., from under the surface. In the main part of the article we explain why this tendency to produce graphene seeds that stand on the surface instead of lying on it is likely to be reduced by Ge monomers diffusing on the surface.

In summary, we find in the initial stage of growth C atoms substitute Ge atoms in surface dimers. At small sub-monolayer coverages, the overwhelming majority of C atoms is expected to occupy this position. As the coverage increases towards a monolayer, more and more C clusters are formed, more and more C substitutional atoms are replaced again by Ge monomers, and more and more C atoms can join the C clusters without previously ejecting a Ge monomer. As a result, carbon agglomerates on top of the substrate surface. Unless a significant amount of Ge monomers becomes trapped in the amorphous network of carbon, the germanium surface should become only slightly roughened, the expected rms being of the order of monatomic step height on on Ge(001), i.e., 0.14\,nm. Any graphene molecules produced in this stage are likely to have many of its edge atoms chemically bonded to the surface Ge atoms and are likely to have a marked tendency to produce 3D structures. This detrimental tendency appears to be opposed by the Ge monomers, which glue graphene edge back to the surface. The diameter of graphene nanocrystallites is expected to be strongly affected by the efficiency of carbon incorporation at sites between carbon edge atoms and germanium surface atoms, while the distribution of the nanocrystallites on the surface is expected to be strongly affected by their diffusion. The latter is likely to be significant only at temperatures above the surface melting point of germanium.

\bibliographystyle{elsarticle-num}
\bibliography{references}

\begin{thebibliography}{10}
\expandafter\ifx\csname url\endcsname\relax
  \def\url#1{\texttt{#1}}\fi
\expandafter\ifx\csname urlprefix\endcsname\relax\def\urlprefix{URL }\fi
\expandafter\ifx\csname href\endcsname\relax
  \def\href#1#2{#2} \def\path#1{#1}\fi

\bibitem{Grapheneroadmap2012}
K.~Novoselov, V.~Falko, L.~Colombo, P.~Gellert, M.~Schwab, K.~Kim, A roadmap
  for graphene, Nature 490 (2012) 192.

\bibitem{reviewferrari2012}
F.~Bonaccorso, A.~Lombardo, T.~Hasan, Z.~Sun, L.~Colombo, A.~Ferrari,
  Production and processing of graphene and 2d crystals, Materials Today 15
  (2012) 564--589.

\bibitem{samsungnature11}
K.~Kim, J.-Y. Choi, T.~Kim, S.-H. Cho, H.-J. Chung, A role for graphene in
  silicon-based semiconductor devices, Nature 479 (2011) 338.

\bibitem{ruoff2012Cu}
L.~Tao, J.~Lee, M.~Holt, H.~Chou, S.~McDonnel, D.~Ferrer, M.~Babenco,
  R.~Wallace, S.~Banerjee, R.~Ruoff, D.~Akinwande, Uniform wafer-scale chemical
  vapor deposition of graphene on evaporated cu(111) film with quality
  comparable to exfoliated monolayer, J. Phys. Chem. C 116 (2012) 24068--24074.

\bibitem{ruoffCVDGraphonCVDBN}
M.~Wang, S.~Jang, W.-J. Jang, M.~Kim, S.-Y. Park, S.-W. Kim, S.-J. Kahng, J.-Y.
  Choi, R.~Ruoff, Y.~Song, S.~Lee, A platform for large-scale graphene
  electronics - cvd growth of single-layer graphene on cvd-grown hexagonal
  boron nitride, Advanced Materials 25 (2013) 2746--2752.

\bibitem{MOS2Grap2012}
Y.~Shi, W.~Zhou, A.-Y. Lu, W.~Fang, Y.-H. Lee, A.~Hsu, S.~Kim, K.~Kim, H.~Yang,
  L.-J. Li, J.-C. Idrobo, J.~Kong, Van der waals epitaxy of mos2 layers using
  graphene as growth templates, Nano Letters 12 (2012) 2784--2791.

\bibitem{IBM2012}
Y.~Wu, K.~Jenkins, A.~Valdes-Garcia, D.~Farmer, Y.~Zhu, A.~Bol,
  C.~Dimitrakopoulos, W.~Zhu, F.~Xia, P.~Avouris, Y.-M. Lin, State-of-the-art
  graphene high-frequency electronics, Nano Letters 12 (2012) 3062--3067.

\bibitem{Sam2012Nanolett}
S.~Vaziri, G.~Lupina, C.~Henkel, A.~Smith, M.~Ostling, J.~Dabrowski,
  G.~Lippert, W.~Mehr, M.~Lemme, A graphene-based hot electron transistor, Nano
  Letters 13 (2013) 1435--1439.

\bibitem{tongayPRX2012}
S.~Tongay, M.~Lemaitre, X.~Miao, B.~Gila, B.~Appleton, A.~Hebard, Rectification
  at graphene-semiconductor interfaces: zero-gap semiconductor-based diodes,
  Physical Review X 2 (2012) 011002.

\bibitem{hackley}
J.~Hackley, D.~Ali, J.~DiPasquale, J.~D. Demaree, C.~J.~K. Richardson,
  Graphitic carbon growth on si(111) using solid source molecular beam epitaxy,
  Applied Physics Letters 95 (2009) 133114.

\bibitem{maeda}
F.~Maeda, H.~Hibino, Study of graphene growth by gas-source molecular beam
  epitaxy using cracked ethanol: Influence of gas flow rate on graphitic
  material deposition, Japanese Journal Applied Physics 50 (2011) 06GE12.

\bibitem{CSiPhasediag}
R.~Olesinski, G.~Abbaschian, The c-si (carbon-silicon) system, Bulletin of
  Alloy Phase Diagrams 5 (1984) 486.

\bibitem{CGePhasediag}
R.~Olesinski, G.~Abbaschian, The c-ge (carbon-germanium) system, Bulletin of
  Alloy Phase Diagrams 5 (1984) 484.

\bibitem{CCuPhasediag}
B.~Predel, \href{http://dx.doi.org/10.1007/10040476_626}{C-cu (carbon-copper)},
  in: O.~Madelung (Ed.), B-Ba -- C-Zr, Vol.~5b of Landolt-Börnstein - Group IV
  Physical Chemistry, Springer Berlin Heidelberg, 1992.
\newblock \href {http://dx.doi.org/10.1007/10040476_626}
  {\path{doi:10.1007/10040476_626}}.
\newline\urlprefix\url{http://dx.doi.org/10.1007/10040476_626}

\bibitem{Mehr2012}
W.~Mehr, J.~Dabrowski, J.~Scheytt, G.~Lippert, Y.~Xie, M.~Lemme, M.~Ostling,
  G.~Lupina, Vertical graphene base transistor, Electron Device Letters 33
  (2012) 691.

\bibitem{GBT_Lecce2013}
V.~D. Lecce, R.~Grassi, A.~Gnudi, E.~Gnani, S.~Reggiani, G.~Baccarani, Graphene
  base transistors: A simulation study of dc and small-signal operation, IEEE
  Transactions on Electron Devices 60 (2013) 3584.

\bibitem{sutter2006Ge}
E.~Sutter, P.~Sutter, Au-induced encapsulation of ge nanowires in protective c
  shells, Advanced Materials 18 (2006) 2583--2588.

\bibitem{angewandt2013}
H.~Kim, Y.~Son, C.~Park, J.~Cho, H.~Choi, Catalyst-free direct growth of a
  single to a few layers of graphene on a germanium nanowire for the anode
  material of a lithium battery, Angew. Chem. Int. Ed. 52 (2013) 5997--6001.

\bibitem{Wang2013srep}
G.~Wang, M.~Zhang, Y.~Zhu, G.~Ding, D.~Jiang, Q.~Guo, S.~Liu, X.~Xie, P.~K.
  Chu, Z.~Di, X.~Wang, Direct growth of graphene film on germanium substrate,
  Scientific Reports 3 (2013) 2465.
\newblock \href {http://dx.doi.org/10.1038/srep02465}
  {\path{doi:10.1038/srep02465}}.

\bibitem{GonBN_Yang2013nmat}
W.~Yang, G.~Chen, Z.~Shi, C.-C. Liu, L.~Zhang, G.~Xie, M.~Cheng, D.~Wang,
  R.~Yang, D.~Shi1, K.~Watanabe, T.~Taniguchi, Y.~Yao, Y.~Zhang, G.~Zhang,
  Epitaxial growth of single-domain graphene on hexagonal boron nitride, Nature
  Materials Letters 12 (2013) 792.

\bibitem{Ge001_surfaceMeltSantoni2003}
a.~V. R.~D. A~Santoni, Electronic structure of the high-temperature ge(1 0 0)
  surface studied by valence band photoemission, Surface Science 537 (2003)
  L423.

\bibitem{Lippert2013Carbon}
G.~Lippert, J.~Dabrowski, Y.~Yamamoto, F.~Herziger, J.~Maultzsch, M.~Lemme,
  W.~Mehr, G.~Lupina, Molecular beam growth of micrometer-size graphene on
  mica, Carbon 52 (2013) 40.

\bibitem{lippert}
G.~Lippert, J.~Dabrowski, M.~Lemme, C.~M. Marcus, O.~Seifarth, G.~Lupina,
  Direct graphene growth on insulator, Physica Status Solidi (b) 248 (2011)
  2619.

\bibitem{Yamamoto2012}
Y.~Yamamoto, G.~Kozlowski, P.~Zaumseil, B.~Tillack, Low threading dislocation
  ge on si by combining deposition and etching, Thin Solid Films 520 (2012)
  3216--3221.

\bibitem{Tuinstra1970}
F.~Tuinstra, J.~L. Koenig, Raman spectrum of graphite, J. Chem. Phys. 53 (1970)
  1126.

\bibitem{thomsen2000}
C.~Thomsen, S.~Reich, Double {R}esonant {R}aman {S}cattering in {G}raphite,
  Physical Review Letters 85 (2000) 5214 -- 5217.

\bibitem{ferrari2006}
A.~C. Ferrari, J.~C. Meyer, V.~Scardaci, C.~Casiraghi, M.~Lazzeri, F.~Mauri,
  S.~Piscanec, D.~Jiang, K.~S. Novoselov, S.~Roth, A.~K. Geim, {R}aman
  {S}pectrum of {G}raphene and {G}raphene {L}ayers, Physical Review Letters 97
  (2006) 187401.

\bibitem{berciaud2009}
S.~Berciaud, S.~Ryu, L.~Brus, T.~Heinz, Probing the {I}ntrinsic {P}roperties of
  {E}xfoliated {G}raphene: {R}aman {S}pectroscopy of {F}ree-{S}tanding
  {M}onolayers, Nano Letters 9~(1) (2009) 346 -- 352.

\bibitem{Kalbac2010}
M.~Kalbac, A.Reina-Cecco, H.~Farhat, J.~Kong, L.~Kavan, M.~Dresselhaus, The
  influence of strong electron and hole doping on the raman intensity of
  chemical vapor deposition graphene, ACS Nano 4 (2010) 6055--6063.

\bibitem{Cheng2010}
Z.~Cheng, Q.~Zhou, C.~Wang, Q.~Li, C.~Wang, Y.~Fang, Toward intrinsic graphene
  surfaces: a systematic study on thermal annealing and wet chemical treatment
  of sio2 supported graphene devices, Nano Letters 11 (2010) 767--771.

\bibitem{Cancado2007}
L.~G. Cançado, A.~Jorio, M.~A. Pimenta, Measuring the absolute raman cross
  section of nanographites as a function of laser energy and crystallite size,
  Phys. Rev. B 76 (2007) 064304.

\bibitem{Lucchese2010}
M.~M. Lucchese, F.~Stavale, E.~H.~M. Ferreira, C.~Vilani, M.~V.~O. Moutinho,
  R.~B. Capaz, C.~A. Achete, , A.~Jorio, Quantifying ion-induced defects and
  raman relaxation length in graphene, Carbon 48 (2010) 1592--1597.

\bibitem{Jeon2006}
C.~Jeon, C.~C. Hwang, T.-H. Kang, K.-J. Kim, B.~Kim, Y.~Chung, C.~Y. Park,
  Evidence from arpes that the ge(001) surface is semiconducting at room
  temperature, Phys. Rev. B 74 (2006) 125407.

\bibitem{Anderson1958}
P.~Anderson, Absence of diffusion in certain random lattices, Phys. Rev. 109
  (1958) 1492--1505.

\bibitem{Peters2012}
E.~C. Peters, A.~J.~M. Giesbers, M.~Burghard, Variable range hopping in
  graphene antidot lattices, Physica Status Solidi (b) 249 (2012) 2522--2525.

\bibitem{Yan2010}
J.~Yan, M.~Fuhrer, Charge transport in dual gated bilayer graphene with corbino
  geometry, Nano Letters 10 (2010) 4521.

\bibitem{Cheah2013}
C.~Y. Cheah, C.~Gómez-Navarro, L.~C. Jaurigue, A.~B. Kaiser, Conductance of
  partially disordered graphene: Crossover from temperature-dependent to
  field-dependent variable-range hopping, arXiv:1305.0315.

\bibitem{Mott1971}
N.~Mott, E.~Davis, Electronic processes in nanocrystalline materials, Oxford
  Univ. Press, London, 1971.

\bibitem{sheetResistanceLi2009}
X.~Li, Y.~Zhu, W.~Cai, M.~Borysiak, B.~Han, D.~Chen, R.~D. Piner, L.~Colombo,
  R.~S. Ruoff, Transfer of large-area graphene films for high-performance
  transparent conductive electrodes, Nano Letters 9 (2009) 4359.

\bibitem{sheetResistanceGunes2010}
F.~G\"unes, H.-J. Shin, C.~Biswas, G.~H. Han, E.~S. Kim, S.~J. Chae, J.-Y.
  Choi, Y.~H. Lee, Layer-by-layer doping of few-layer graphene film, ACS Nano 4
  (2010) 4595.

\bibitem{sheetResistanceBiswas2011}
C.~Biswas, Y.~H. Lee, Graphene versus carbon nanotubes in electronic devices,
  Advanced Functional Materials 21 (2011) 3806.

\bibitem{sheetResistanceIHPremark}
In our experiments, sheet resistance of CVD graphene transferred from Cu to
  \ce{SiO2}/Si is typically between 1 and 4\,k$\Omega/\Box$.

\bibitem{OswaldLifshitz1961}
I.~Lifshitz, V.~Slyozov, The kinetics of precipitation from supersaturated
  solid solutions, Journal of Physics and Chemistry of Solids 19 (1961) 35.

\bibitem{Nieminen2003}
P.~Lehtinen, A.~S. Foster, A.~Ayuela, A.~Krasheninnikov, K.~Nordlund, ,
  R.~Nieminen, Magnetic properties and diffusion of adatoms on a graphene
  sheet, Physical Review Letters 91 (2003) 017202.

\bibitem{nanoribbon_Zhang2009}
Z.~Zhang, W.~Guo, Electronic properties of zigzag graphene nanoribbons on
  si(001), Applied Physics Letters 95 (2009) 023107.

\bibitem{SiinGeDiffusionRaisanen1891}
J.~R\"ais\"anen, J.~Hirvonen, A.~Anttila,
  \href{http://www.sciencedirect.com/science/article/pii/0038110181900277}{The
  diffusion of silicon in germanium}, Solid-State Electronics 24~(4) (1981)
  333.
\newblock \href
  {http://dx.doi.org/http://dx.doi.org/10.1016/0038-1101(81)90027-7}
  {\path{doi:http://dx.doi.org/10.1016/0038-1101(81)90027-7}}.
\newline\urlprefix\url{http://www.sciencedirect.com/science/article/pii/0038110181900277}

\bibitem{Yamamoto2011}
Y.~Yamamoto, P.~Zaumseil, T.~Arguirov, M.~Kittler, B.~Tillack, Low threading
  dislocation density ge deposited on si (100) using rpcvd, Solid State
  Electronics 60 (2011) 2--6.

\bibitem{Klesse2011Capellini}
W.~Klesse, G.~Scappucci, G.~Capellini, M.~Simmons, Preparation of the ge(001)
  surface towards fabrication of atomic scale germanium devices, Nanotechnology
  22 (2011) 145604.

\bibitem{ConGstickingPhilips}
V.~Philipps, E.~Vietzke, K.~Flaskamp, Sticking probabilities of evaporated
  \ce{C1}, \ce{C2}, and \ce{C3} on pyrolitic graphite, Surface Science 178
  (1986) 806.

\bibitem{quantespresso}
P.~Gianozzi, Quantum espresso: a modular and open-source software project for
  quantum simulations of materials, Journal of Physics: Condensed Matter 21
  (2009) 395502.

\bibitem{PBE}
J.~P. Perdew, K.~Burke, M.~Ernzerhof, Generalized gradient approximation made
  simple, Physical Review Letters 77 (1996) 3865--3868.

\bibitem{NEB}
H.~Jonsson, G.~Mills, K.~W. Jacobsen, Nudged elastic band method for finding
  minimum energy paths of transitions, in: B.~J. Berne, G.~Ciccotti, D.~F. Coke
  (Eds.), Classical and Quantum Dynamics in Condensed Phase Simulations, World
  Scientific, 1998, p. 385.

\bibitem{NEB_CI}
G.~Henkelman, B.~Uberuaga, H.~Jonsson, A climbing image nudged elastic band
  method for finding saddle points and minimum energy paths, Journal of
  Chemical Physics 113 (2000) 9901--9904.

\bibitem{Molle2006}
A.~Molle, M.~Bhuiyan, G.~Tallarida, M.~Fanciulli, In situ chemical and
  structural investigations of the oxidation of ge(001) substrates by atomic
  oxygen, Applied Physics Letters 89 (2006) 083504.

\bibitem{sun2006}
S.~Sun, Y.~Sun, Z.~Liu, D.-I. Lee, S.~Peterson, P.~Pianetta, Surface
  termination and roughness of ge(100) cleaned by hf and hcl solutions, Applied
  Physics Letters 88 (2006) 021903.

\bibitem{landoltSiC}
SGTE, P.~Franke, D.~N. (Ed.), Binary systems c-si, in Landolt-Boernstein:
  Numerical data and relationships in science and technology IV/19B2 (2004)
  140--142.

\bibitem{xiaolinLi}
X.~Li, G.~Zhang, X.~Bai, X.~Sun, X.~Wang, E.~Wang, H.~Dai, Highly conducting
  graphene sheets and langmuir- blodgett films, Nature Nanotechnology 3 (2008)
  538--542.

\bibitem{XPS11}
E.~Moreau, S.~Godey, X.~Wallart, I.~Razado-Colambo, J.~Avila, M.-C. Asensio, ,
  D.~Vignaud, High-resolution angle-resolved photoemission spectroscopy study
  of monolayer and bilayer graphene on the c-face of sic, Physical Review B 88
  (2013) 075406.

\bibitem{Reich2004}
S.~Reich, C.~Thomsen, Raman spectroscopy of graphite, Phil. Trans. R. Soc.
  Lond. A 362 (2004) 2271--2288.

\bibitem{Ferrari2000}
A.~Ferrari, J.~Robertson, Interpretation of raman spectra of disordered and
  amorphous carbon, Phys. Rev. B 61 (2000) 14095.

\bibitem{Reina2009}
A.~Reina, X.~Jia, J.~Ho, D.~Nezich, H.~Son, V.~Bulovic, M.~S. Dresselhaus,
  J.~Kong, Large area, few-layer graphene films on arbitrary substrates by
  chemical vapor deposition, Nano Letters 9 (2009) 30--35.

\bibitem{Oliveira2013}
M.~O. Jr., T.~Schumann, R.~Gargallo-Caballero, F.~Fromm, T.~Seyller,
  M.~Ramsteiner, A.~Trampert, L.~Geelhaar, J.~Lopes, H.~Riechert, Mono- and
  few-layer nanocrystalline graphene grown on al2o3(0001) by molecular beam
  epitaxy, Carbon 56 (2013) 339.

\bibitem{Poncharal2008}
P.~Poncharal, A.~Ayari, T.~Michel, J.-L. Sauvajol, Raman spectra of misoriented
  bilayer graphene, Phys. Rev. B 78 (2008) 113407.

\bibitem{Herziger2012}
F.~Herziger, P.~May, J.~Maultzsch, Layer-number determination in graphene by
  out-of-plane phonons, Phys. Rev. B 85 (2012) 235447.

\bibitem{Lui2013}
C.~Lui, T.~Heinz, Measurement of layer breathing mode vibrations in few-layer
  graphene, Phys. Rev. B 87 (2013) 121404R.

\bibitem{ReichBook2004}
S.~Reich, C.~Thomsen, J.~Maultzsch, Carbon nanotubes: basic concepts and
  physical properties, Wiley-VCH, Weinheim, 2004.

\bibitem{Schroder1998}
D.~Schroder, Semiconductor material and device characterization, J. Wiley and
  Sons, New York, 1998.

\bibitem{Kevan1985}
S.~D. Kevan, Surface states and reconstruction on ge(001), Phys. Rev. B 32
  (1985) 2344--2350.

\bibitem{Eriksson2008}
P.~E.~J. Eriksson, M.~Adell, K.~Sakamoto, R.~I.~G. Uhrberg, Origin of a surface
  state above the fermi level on ge(001) and si(001) studied by
  temperature-dependent arpes and leed, Phys. Rev. B 77 (2008) 085406.

\bibitem{CinGsolubilityHaller1982}
E.~E. Haller, W.~L. Hansen, P.~Luke, R.~McMurray, B.~Jarret, Carbon in high
  purity germanium, IEEE Transactions on Nuclear Science 29 (1982) 745.

\bibitem{Pedersen1999}
T.~P.~L. Pedersen, A.~N. Larsen, A.~Mesli, Carbon-related defects in
  proton-irradiated, n-type epitaxial si$_{1-x}$ge$_x$, Applied Physics Letters
  75 (1999) 4085.

\end{thebibliography}

\end{document}